\documentclass{aa}
\usepackage{natbib}

\usepackage{amsmath,amssymb}
\usepackage[varg]{txfonts}

\usepackage{color}
\usepackage{graphicx}

\usepackage{epstopdf}

\usepackage{gensymb}
\usepackage{stfloats}

\usepackage[utf8]{inputenc}
\usepackage{hyperref}
\usepackage{booktabs,multirow}

\usepackage{tablefootnote}

\usepackage{ulem}
\definecolor{mygray}{gray}{0.6}
\definecolor{darkblue}{rgb}{0.0, 0.0, 0.8}

\newcommand{\se}[1]{Sect.~\ref{sec:#1}}
\newcommand{\eq}[1]{Eq.~(\ref{eq:#1})}

\newcommand{\fg}[1]{Fig.~\ref{fig:#1}}
\newcommand{\Fg}[1]{Figure~\ref{fig:#1}}
\newcommand{\tb}[1]{Table~\ref{tab:#1}}

\begin{document}

\title{Pebble-driven planet formation for TRAPPIST-1 and other compact systems}
\author{Djoeke Schoonenberg\inst{1}, Beibei Liu\inst{2}, Chris W.~Ormel\inst{1}, and Caroline Dorn\inst{3}}
\authorrunning{D.~Schoonenberg et al.}
\institute{Anton Pannekoek Institute for Astronomy, University of Amsterdam, Science Park 904, 1090 GE Amsterdam, The Netherlands\label{inst1} \\
\email{d.schoonenberg@uva.nl}
\and
Lund Observatory, Department of Astronomy and Theoretical Physics, Lund University, Box 43, 221 00 Lund, Sweden\label{inst2}
\and
University of Zurich, Institute of Computational Sciences, Winterthurerstrasse 190, CH-8057, Zurich, Switzerland\label{inst3}}
\date{\today}

\abstract{Recently, seven Earth-sized planets were discovered around the M-dwarf star TRAPPIST-1. Thanks to transit-timing variations, the masses and therefore the bulk densities of the planets have been constrained, suggesting that all TRAPPIST-1 planets are consistent with water mass fractions on the order of 10\%. These water fractions, as well as the similar planet masses within the system, constitute strong constraints on the origins of the TRAPPIST-1 system. In a previous work, we outlined a pebble-driven formation scenario. In this paper we investigate this formation scenario in more detail. We used a Lagrangian smooth-particle method to model the growth and drift of pebbles and the conversion of pebbles to planetesimals through the streaming instability. We used the N-body code \texttt{MERCURY} to follow the composition of planetesimals as they grow into protoplanets by merging and accreting pebbles. This code is adapted to account for pebble accretion, type-I migration, and gas drag. In this way, we modelled the entire planet formation process (pertaining to planet masses and compositions, not dynamical configuration). We find that planetesimals form in a single, early phase of streaming instability. The initially narrow annulus of planetesimals outside the snowline quickly broadens due to scattering. Our simulation results confirm that this formation pathway indeed leads to similarly-sized planets and is highly efficient in turning pebbles into planets. Our results suggest that the innermost planets in the TRAPPIST-1 system grew mostly by planetesimal accretion at an early time, whereas the outermost planets were initially scattered outwards and grew mostly by pebble accretion. The water content of planets resulting from our simulations is on the order of 10\%, and our results predict a `V-shaped' trend in the planet water fraction with orbital distance: from relatively high (innermost planets) to relatively low (intermediate planets) to relatively high (outermost planets).}


\maketitle
\section{Introduction}
The cool dwarf star TRAPPIST-1 has been found to be the host of an ultra-compact system of seven Earth-sized planets \citep{2016Natur.533..221G, 2017Natur.542..456G, 2017NatAs...1E.129L}. All seven planets are similar in size (about an Earth radius), which is puzzling to explain with classical planet formation theories \citep{OLS2017}. Moreover, measurements of transit-timing variations and dynamical modelling have constrained their bulk densities, and it has been shown that these densities are consistent with water fractions of a few to tens of mass percent \citep{2018A&A...613A..68G,2018RNAAS...2c.116U,2018ApJ...865...20D}. This poses another challenge to planet formation theory. If the planets have formed interior to the water snowline (the distance from the star beyond which water condenses to solid ice), one would expect their water content to be much lower; on the other hand, if they had formed outside of the snowline, one would expect even higher water fractions.

\citet{OLS2017} proposed a scenario for the formation of a compact planetary system like TRAPPIST-1, assuming an inward flux of icy pebbles from the outer disk. In this scenario, planetesimals form in a narrow ring outside the snowline, due to a local enhancement in solids that triggers the streaming instability. The solids enhancement materialises because of outward diffusion of water vapour and condensation \citep{1988Icar...75..146S,2006Icar..181..178C,2013A&A...552A.137R,SO2017,2017A&A...608A..92D}. The resulting icy planetesimals merge and accrete pebbles until they are large enough to start migrating towards the star by type-I migration. Once the protoplanet has migrated across the snowline, the pebbles it accretes are water-poor. The total mass in icy material a protoplanet accretes outside the snowline versus the total mass in dry material it accretes inside the snowline determines its eventual bulk composition. In the case of the TRAPPIST-1 system, this process repeats itself until seven planets are formed, such that the snowline region acts as a `planet formation factory'. In this formation model, the similar masses are a result of planet growth stalling at the pebble isolation mass, and the moderate water contents are a result of the combination of wet and dry accretion. \citet{2018NatAs...2..297U} have also proposed a formation model for TRAPPIST-1 where inward migration is key to explaining the moderate water fractions, however in their model planet assembly processes are not included.

In this paper we present a numerical follow-up study of \citet{OLS2017}. Dust evolution and planetesimal formation are treated with the Lagrangian smooth-particle method presented in \citet{2018A&A...620A.134S}. The pebble flux from the outer disk is no longer a free parameter as in \citet{OLS2017}, but follows from the dust evolution code. This code is coupled to the N-body code \texttt{MERCURY} \citep{1999MNRAS.304..793C}, which is adapted to account for pebble accretion, gas drag, and type-I migration \citep{2019arXiv190210062L}. Our model does not follow the later dynamical evolution -- the re-arrangement of the planetary system architecture during disk dissipation \citep{2017A&A...601A..15L}, or the characteristics and long-term stability of the resonant chain \citep{2018MNRAS.476.5032P,2017MNRAS.470.1750I,2019arXiv190208772I} -- but it does follow the entire planet formation process from small dust particles to full-sized planets, whilst keeping track of the solid bodies' water content. Any model that aims to explain the origin of the TRAPPIST-1 planets must, of course, be able to explain their observed masses and compositions.

In \se{models}, we summarise the features of the codes that we employ in this work and describe how they are coupled. We present results of our fiducial model runs in \se{results} and describe how the results depend on parameter choices in \se{parameters}. We discuss our results in \se{discussion} and summarise our main findings in \se{conclusions}.

\section{Model}\label{sec:models}
\subsection{Disk model}\label{sec:disk}
The gas disk in which planet formation takes place is modelled as a one-dimensional axisymmetric disk, assuming a viscously-relaxed (steady-state) gas surface density profile $\Sigma_{\rm{gas}}$:
\begin{equation}\label{eq:sigma}
\Sigma_{\rm{gas}} = \frac{\dot{M}_{\rm{gas}}}{3 \pi \nu},
\end{equation}
where $\dot{M}_{\rm{gas}}$ is the gas accretion rate, and the viscosity $\nu$ is related to the dimensionless turbulence parameter $\alpha$ \citep{1973A&A....24..337S,1974MNRAS.168..603L} as
\begin{equation}
\nu = \alpha c_{s}^{2} \Omega^{-1},
\end{equation}
with $c_{s}$ the sound speed and $\Omega$ the Keplerian orbital frequency. The sound speed $c_{s}$ is given by
\begin{equation}
c_{s} = \sqrt{\frac{k_{B} T (r)}{\mu}},
\end{equation}
where $k_{B}$ is the Boltzmann constant and $\mu$ is the mean molecular weight of the protoplanetary disk gas, which we set to 2.34 times the proton mass. Concerning the temperature $T$ as a function of radial distance to the star $r$, we consider viscous heating and stellar irradiation, and define $T(r)$ as
\begin{equation}
T(r) = [T_{\rm{visc}}^{4} (r) + T_{\rm{irr}}^{4} (r)]^{1/4} ,
\end{equation}
where $T_{\rm{visc}} (r)$ is the viscous temperature profile, which in our standard model is given by
\begin{equation}\label{eq:vis}
T_{\rm{visc}} (r) = 180 \left( \frac{r}{0.1 \: \rm{au}}\right)^{-1} \rm{K},
\end{equation}
and $T_{\rm{irr}} (r)$ is the irradiation temperature profile, given by
\begin{equation}
T_{\rm{irr}} (r) = 150 \left( \frac{r}{0.1 \: \rm{au}}\right)^{-1/2} \rm{K}.
\end{equation}
We assume a vertically isothermal disk, leading to a disk scale height $H_{\rm{gas}}$ of
\begin{equation}
H_{\rm{gas}} = c_{s} / \Omega.
\end{equation}
The gas moves with a velocity $v_{\rm{gas}} = - 3\nu / 2r$. 

\subsection{Dust evolution and planetesimal formation}
The evolution of dust and formation of planetesimals are treated with a Lagrangian smooth-particle method. Here, we only give a summary of the characteristics of this model. For more information we refer to \citet{2018A&A...620A.134S}. 

\subsubsection{Dust evolution}
We assume that the dust surface density $\Sigma_{\rm{dust}}$ is initially a radially constant fraction of the gas surface density $\Sigma_{\rm{gas}}$. This radially constant fraction is called the metallicity $Z$, which we set to 2\%. Outside of the water snowline, grains consist of 50\% water ice and 50\% refractory material; interior to the water snowline, the dust composition is completely refractory. Dust grains all start out with the same size (0.1 $\mu$m). At any time, the dust grain size distribution at a given distance from the star is mono-disperse: particles are described by a single size at a given time point and location in the disk. Dust grains grow collisionally assuming perfect sticking \citep{2016A&A...586A..20K}, and fragment when their mutual impact velocities become larger than the fragmentation threshold velocity, which we set to 10 m/s for ice-coated particles and 3 m/s for refractory particles, motivated by laboratory experiments \citep{1993Icar..106..151B, 1999A&A...347..720S, 2015ApJ...798...34G}. Although it has recently been reported that the sticking properties of ice vary with temperature \citep{2019ApJ...873...58M}, for simplicity we keep the fragmentation thresholds constant with the semi-major axis.

Initially, when the dust grains are small, the dust is vertically distributed in the same way as the gas; the solids scale height $H_{\rm{solids}}$ is initially equal to the gas scale height $H_{\rm{gas}}$. When dust grains have grown large enough to start decoupling aerodynamically from the surrounding gas, however, they settle to the disk midplane. Vertical settling results in a solids scale height that is smaller than the gas scale height \citep{2007Icar..192..588Y}:
\begin{equation}
H_{\rm{solids}} = \sqrt{\frac{\alpha}{\tau + \alpha}} H_{\rm{gas}},
\end{equation}
where $\tau$ is the dimensionless stopping time of the particles. In calculating $\tau$, we take into account the composition (water fraction) of the particles, and treat both the Epstein and the Stokes drag regimes. Particles with a non-negligible dimensionless stopping time ($\tau \gtrsim 10^{-3}$) are called pebbles\footnote{The solids surface density $\Sigma_{\rm{dust}}$ covers all solid particles. Pebbles belong to the same class of particles as dust grains, which have much smaller stopping times.}. Besides settling vertically, pebbles also move radially due to angular momentum loss as a result of experiencing a headwind from the gas. Taking into account the back-reaction of the solid particles onto the gas, the radial velocity of solid particles is given by \citep{NakagawaEtal1986}
\begin{equation}
v_{\rm{p}} = - \frac{2 \eta v_{K} \tau - v_{\rm{gas}} (1 + \xi)}{(1 + \xi)^{2} + \tau^{2}},
\end{equation}
where $\xi$ is the midplane solids-to-gas volume density ratio, $v_{K}$ the Keplerian velocity, and $\eta$ the `sub-Keplerianity' or headwind factor,
\begin{equation}
\eta v_{K} = - \frac{1}{2} \frac{c_{s}^{2}}{v_K}\frac{\partial \log P}{\partial \log r},
\end{equation}
where $P$ is the gas pressure.
The radial pebble flux $\dot{M}_{\rm{peb}}$ is calculated by
\begin{equation}
\dot{M}_{\rm{peb}} = \Sigma_{\rm{dust}} 2 \pi r |v_{\rm{p}}|.
\end{equation}
The location of the water snowline depends on the temperature structure, as well as on the water vapour pressure - which in turn depends on the flux of icy pebbles from the outer disk that deliver water vapour to the inner disk (e.g. \citet{2006Icar..181..178C,2015arXiv151105563P, SO2017}). We therefore evaluate $\dot{M}_{\rm{peb}}$ outside the water snowline to calculate the position of the water snowline (for more details, see \citet{SO2017, 2018A&A...620A.134S}).

\begin{figure*}
        \centering
                \includegraphics[width=\textwidth]{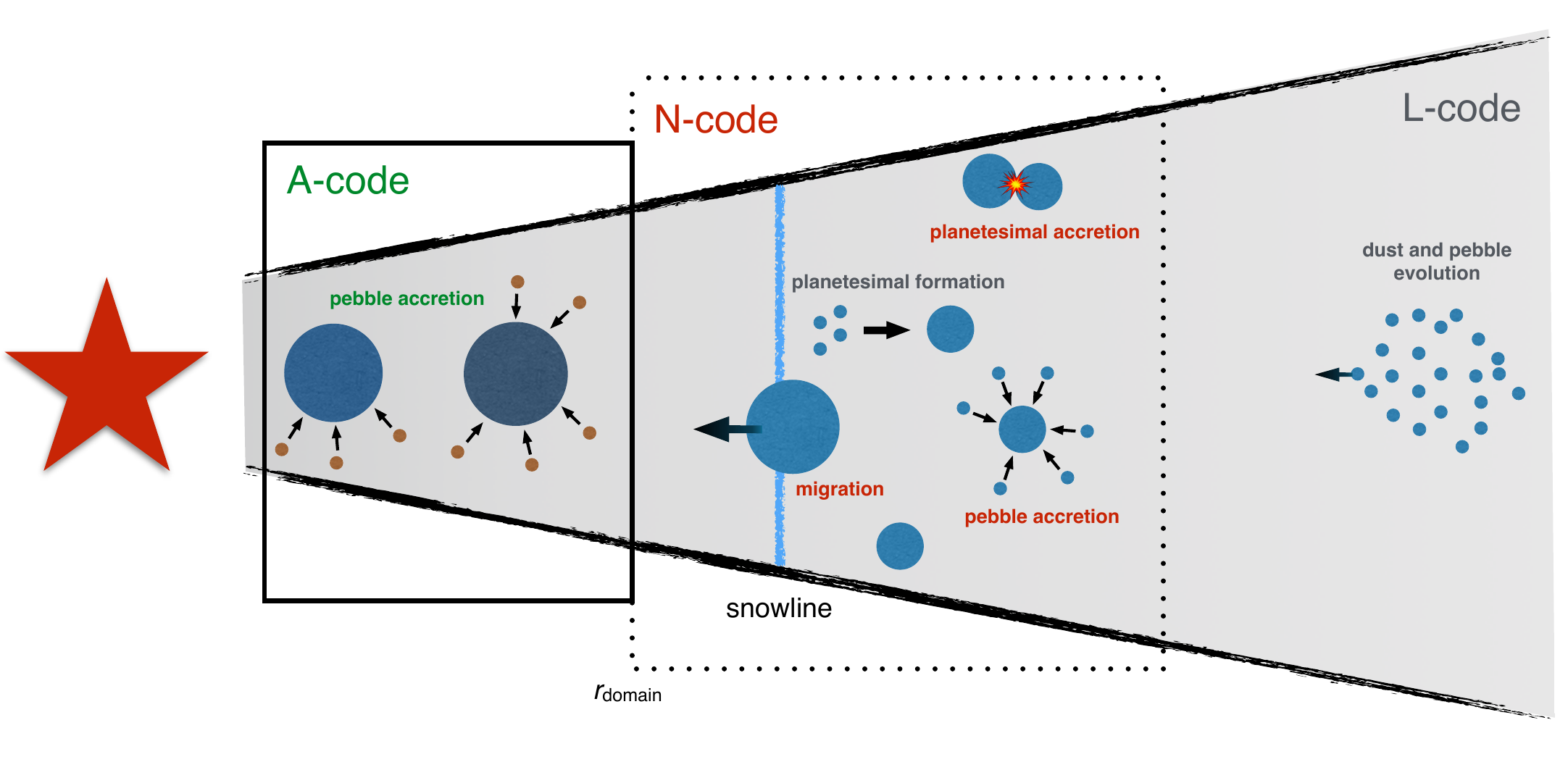}
\caption{Cartoon of the code construction used to simulate the planet formation process from dust grains to full-sized planets. The evolution of dust and pebbles and planetesimal formation are treated by the Lagrangian code (`L-code'), which covers the entire disk. Planetesimals form just outside the snowline, and their growth and migration are followed with an N-body code (`N-code'), which accounts for dynamics, planetesimal and pebble accretion, gas drag, and type-I migration. When protoplanets migrate across the inner domain radius of the N-code $r_{\rm{domain}}$, their final growth stage is calculated with a semi-analytic model (`A-code'). The way in which the codes are coupled in time is discussed in \se{couple}.}\label{fig:completecartoon}
\end{figure*}

\subsubsection{Planetesimal formation}
The pebble mass flux outside the water snowline also regulates planetesimal formation outside the snowline. The combination of a strong gradient in the water vapour distribution across the water snowline together with some degree of turbulence leads to an outward diffusive flux of water vapour \citep{1988Icar...75..146S,2006Icar..181..178C,2013A&A...552A.137R,SO2017,2017A&A...608A..92D}. The vapour that has been transported outwards condenses onto inward-drifting icy pebbles, leading to a locally enhanced solids-to-gas ratio. This enhancement can be large enough to reach the conditions for streaming instability, such that planetesimals can form in an annulus outside the snowline. We have quantified the critical pebble flux $\dot{M}_{\rm{peb, crit}}$ needed to trigger streaming instability as a function of $\tau$, $\alpha,$ and $\dot{M}_{\rm{gas}}$ \citep{SO2017, 2018A&A...620A.134S}. If the pebble flux measured outside the water snowline exceeds the critical value $\dot{M}_{\rm{peb, crit}}$ for a given set of disk conditions, planetesimals form at a rate $d M_{\rm{pltsml}} / d t = \dot{M}_{\rm{peb}} - \dot{M}_{\rm{peb, crit}}$; the excess pebble flux is transformed to planetesimals. More generally, at any time
\begin{equation}\label{eq:excess}
\frac{d M_{\rm{pltsml}}}{d t} = \max [\dot{M}_{\rm{peb}} - \dot{M}_{\rm{peb, crit}}, 0].
\end{equation}
We assume that all formed planetesimals have the same initial size. In our fiducial model, this initial size is 1200 km, corresponding to a mass of $1.8 \times 10^{-3}$ Earth masses (for a planetesimal internal density of 1.5 $\rm{g} \: \rm{cm}^{-3}$, corresponding to a composition of 50\% ice and 50\% silicates \citep{2018A&A...620A.134S}). Simulations show that the initial mass function of planetesimals formed by streaming instability can be described by a power law (possibly with an exponential cutoff), such that the typical size of formed planetesimals is a few hundred kilometers in size \citep{2015SciA....1E0109J,2016ApJ...822...55S,2017A&A...597A..69S,2018arXiv181010018A}, although there is considerable variation amongst these simulations, related to the choice of parameters such as simulation box size and disk metallicity. Our initial planetesimal size of 1200 km is a few factors larger. However, smaller initial planetesimal sizes lead to a larger number of planetesimals and therefore to longer computational timescales. For computational reasons, therefore, we have chosen this value\footnote{With this initial size, $\sim$800 planetesimals are formed in our fiducial model. A single simulation takes approximately a week on a modern desktop computer.}. We have verified that our results do not depend greatly on the initial planetesimal size (\se{parameters}). Practically, an initial planetesimal size of 1200 km means that a single planetesimal forms every time enough pebble mass has been transformed to planetesimal mass to materialise a new 1200 km-sized planetesimal. The location where this planetesimal forms is assumed to be the location where the solids-to-gas ratio outside the snowline peaks, according to the results of \citet{SO2017}. Planetesimals therefore form in a narrow annulus just exterior to the snowline.

\subsection{Planetesimal growth and migration}\label{sec:mercury}
The planetesimals that form in the Lagrangian dust evolution code described above are followed with an adapted version of the \texttt{MERCURY} N-body code \citep{1999MNRAS.304..793C,2019arXiv190210062L}. In this code, planetesimals can grow by planetesimal accretion as well as by pebble accretion. Moreover, gas drag and \hbox{type-I} migration are taken into account. More details can be found in \citet{2019arXiv190210062L}. For planetesimals with the initial size of 1200 km, the migration timescale is still very long. Only after planetesimals have grown larger (by accreting pebbles and other planetesimals) do they start to migrate inwards. Computational times become longer as planetesimals migrate inwards and their orbital periods decrease. To reduce the computational time we therefore take protoplanets out of the N-body code when they cross some (quite arbitrary) distance $r_{\rm{domain}}$. We take $r_{\rm{domain}} = 0.7 r_{\rm{snow}}$ to ensure that the removed bodies are in the water-free growth regime (see \se{wet} for further discussion). The last growth phase of protoplanets, which occurs after they have crossed $r_{\rm{domain}}$, is treated with a semi-analytic model and is discussed in \se{last}.

We do not take into account the effects on the disk structure of a rapid opacity variation across the snowline -- due to the abundance of small grains released by evaporating icy pebbles -- leading to a pressure bump (e.g. \citet{2007ApJ...664L..55K,2014A&A...570A..75B}). A pressure bump -- or more generally a local flattening of the pressure profile -- would lead to a pile-up of pebbles, thereby aiding planetesimal formation at the snowline. The formation of more icy planetesimals would result in higher planet water fractions, since planet growth by planetesimal accretion would be more efficient. The effects of an opacity transition on the migration rate remains unclear, but one possibility is that the snowline acts as a migration trap (e.g. \citet{2014A&A...570A..75B,2019MNRAS.484..345C}). If that were the case, our planet formation model that is based on the inward migration of protoplanets across the snowline would not hold. However, it must be appreciated that these effects heavily depend on the exact thermodynamic state of the disk, which carries major uncertainties (e.g. the efficiency of the magneto-rotational instability and the rate at which micron-sized grains coagulate to lower the opacity). For simplicity, therefore, we have sidelined these subtleties and assumed a simple disk structure.

\begin{figure}
        \centering
                \includegraphics[width=0.49\textwidth]{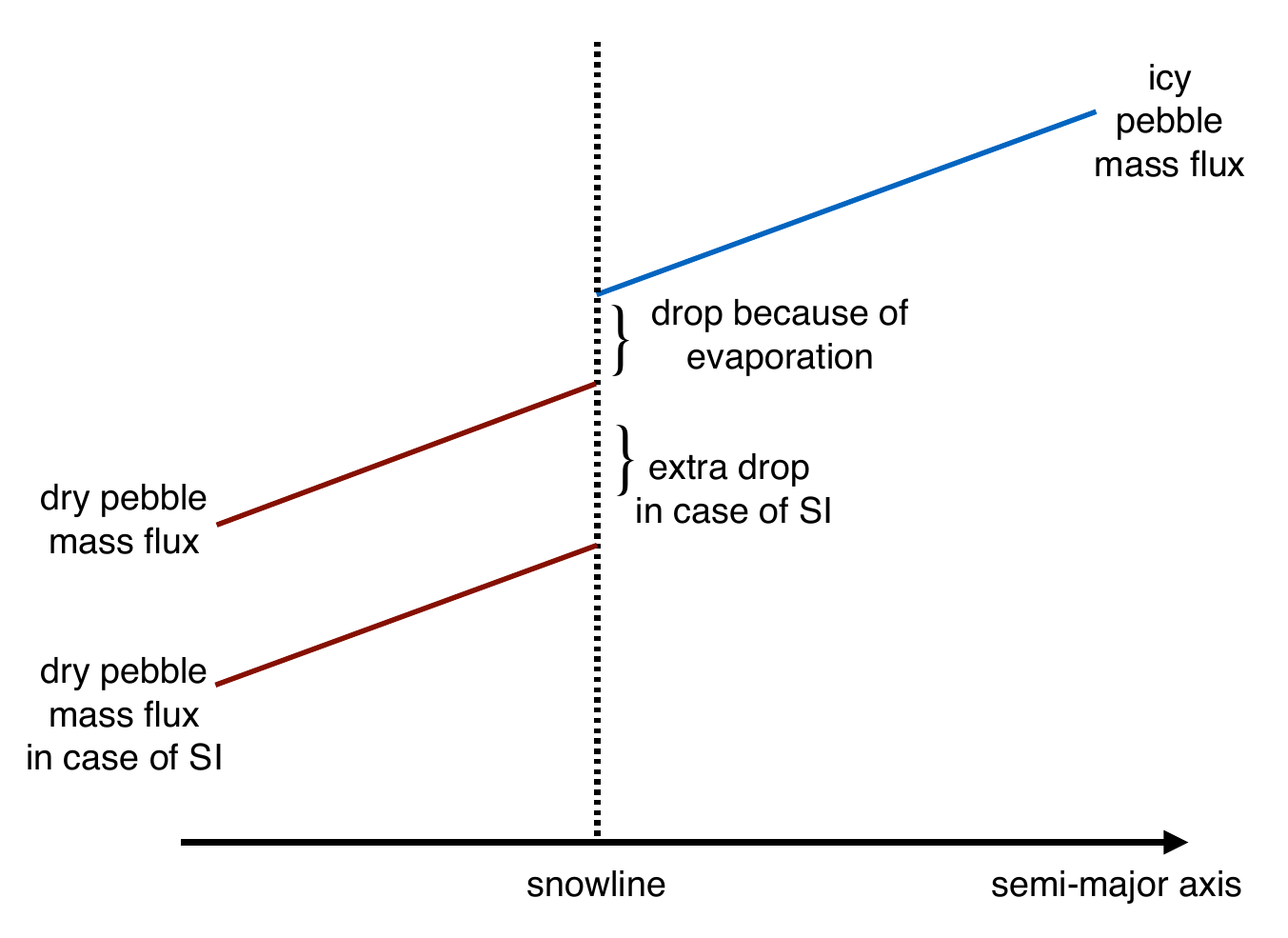}
\caption{Sketch of the pebble mass flux as function of semi-major axis in the N-body code. The pebble mass flux decreases with decreasing semi-major axis because of pebble accretion. Interior to the snowline, the pebble mass flux is halved due to evaporation. When streaming instability (SI) is active, the pebble mass flux at the snowline is decreased even further due to planetesimal formation, as dictated by the Lagrangian dust code.}\label{fig:pebbleflux_cartoon}
\end{figure}

\subsection{Coupling of the two codes}\label{sec:couple}
We couple the Lagrangian dust evolution code (`L-code'), the N-body code (`N-code'), and the semi-analytic `last growth phase' code (`A-code') in a self-consistent way, enabling us to model the entire planet assembly process (we do not model the long-term dynamical evolution of the planets). The gas disk model (\se{disk}) is the same in all codes. \Fg{completecartoon} provides a cartoon of the simulation setup. The L-code covers the entire disk and describes the evolution of dust and pebbles and planetesimal formation outside the water snowline, as well as the evolution of dry (silicate) pebbles interior to the snowline. The dynamics and growth of planetesimals (by planetesimal and pebble accretion) are followed by the N-code. Interior to $r_{\rm{domain}}$, the growth of protoplanets by dry pebble accretion is treated with a semi-analytic model (`A-code'), described in more detail in \se{last}. In this section we describe how the L-code and N-code are coupled.

Information about the pebbles (the pebble flux, as well as the size and composition of pebbles as a function of distance to the star) is provided by the L-code to the N-code. When streaming instability is taking place, the pebble flux that reaches interior to the snowline is reduced because part of the pebble flux outside the water snowline (the excess flux; $\dot{M}_{\rm{peb}} - \dot{M}_{\rm{peb, crit}}$) is converted to planetesimals. Therefore, during streaming instability, the pebble flux reaching inside of the water snowline is given by 0.5 $\dot{M}_{\rm{peb, crit}}$, where the factor 0.5 takes into account the evaporation of the pebbles' water content. This is sketched in \fg{pebbleflux_cartoon}. Pebbles drift in from the outer disk and are accreted by planetesimals just outside the snowline. The pebble flux therefore decreases with decreasing semi-major axis.

The formation times and locations of planetesimals are provided for the N-code as well. The formed planetesimals are injected into the N-code at their formation times. The locations where planetesimals are injected are picked randomly within an annulus centred on the formation location dictated by the L-code, but with a finite width of $\eta r, $  which is the typical length scale of streaming instability filaments as found by simulations \citep{2016ApJS..224...39Y,2018ApJ...862...14L}. The eccentricities and inclinations of the planetesimals are initialised according to a Rayleigh distribution \citep{2019arXiv190210062L}. When planetesimals (or proto-planets) migrate across the snowline, the characteristics of the pebbles they can accrete change (typically, in our simulations, icy pebbles outside the snowline have dimensionless stopping times $\tau \sim 10^{-1}$, corresponding to a physical size of a few centimetres, and silicate grains interior to the snowline have $\tau \sim 10^{-3}$, corresponding to physical sizes of less than a millimetre); this information is provided to the N-code by the L-code.

\begin{figure*}[t]
        \centering
                \includegraphics[width=0.9\textwidth]{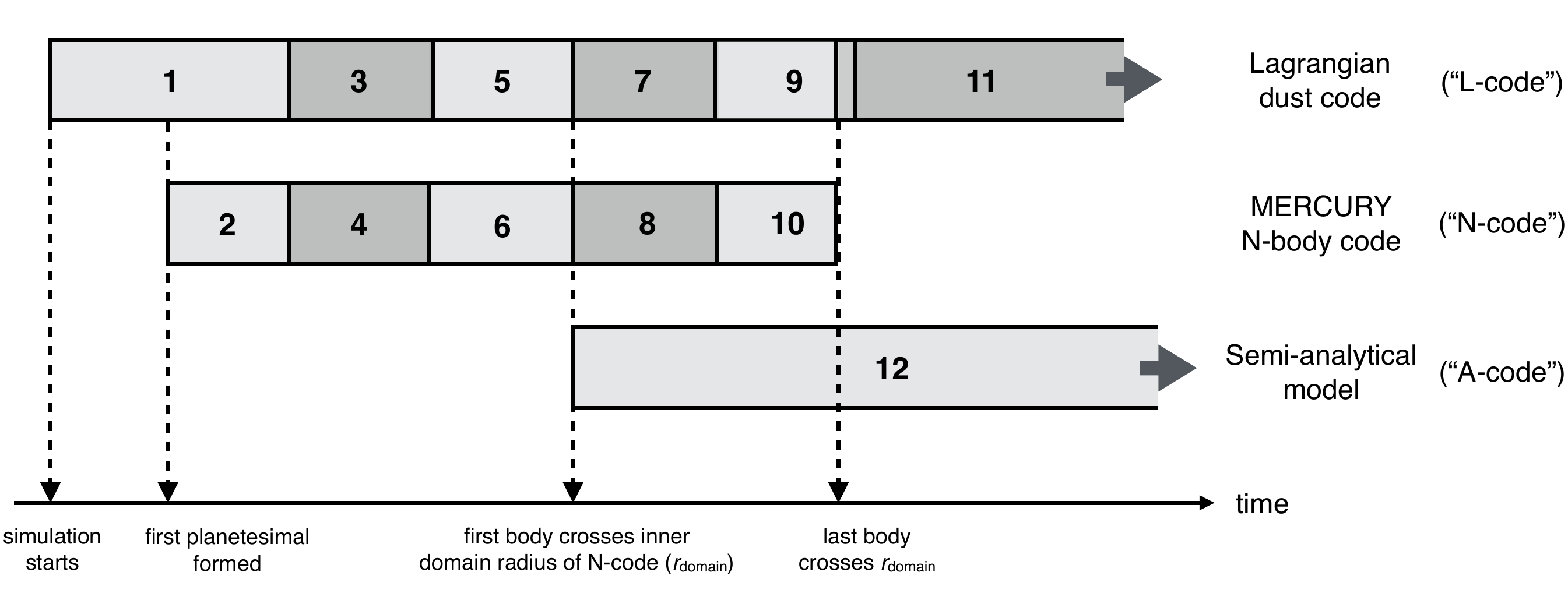}
\caption{Sketch of the code coupling. The code runs in blocks, starting at block 1. For illustration purposes the total number of blocks is 12 in this sketch; in reality, there are many more blocks in one simulation. After some planetesimals have formed during the first run of the Lagrangian dust evolution code (top row), we switch to the N-body code and inject the formed planetesimals (middle row). After a fixed time interval we switch back to the Lagrangian dust code and update the value for the pebble accretion efficiency as was output by the N-body code. This `zig-zag' behaviour continues until all planetesimals have left the N-body code, either because they have been accreted by bigger objects, or because they have grown into bigger objects and have migrated across the inner radius of the N-body domain $r_{\rm{domain}}$ (see text for more details). The final stage of our calculation concerns the growth of the protoplanets that have crossed $r_{\rm{domain}}$, and is discussed in \se{last}.}\label{fig:coupling}
\end{figure*}

The pebble accretion efficiency of the planetesimal population outside of the snowline that follows from the N-code is communicated back to the L-code. Here we assume that the fraction of pebbles that are accreted by the already existing planetesimals outside the snowline, as found by the N-code, are not available to form new planetesimals in the L-code. In other words, anywhere exterior to the snowline, pebble accretion happens before planetesimal formation.

The mutual feedback between the two codes compels us to run them in a `zig-zag' way. To avoid any biases in the code outputs, we switch from one code to the other after fixed time intervals. The length of these time intervals is chosen such that the output values (e.g. pebble mass flux, pebble accretion efficiency) do not change significantly, but that a significant number of planetesimals are still formed within an interval. We choose a time interval of 5000 years, which meets both these requirements. The coupling of the two codes is illustrated in \fg{coupling}. The top row depicts the Lagrangian dust evolution and planetesimal formation code; the middle row depicts the N-body code. The code starts at the block indicated by the number 1, then moves to block 2, and so on. The bottom row corresponds to the semi-analytic part of the code that deals with the last phase of planet growth, after planetesimals have crossed the inner domain radius of the N-body code. We will discuss this semi-analytic calculation in the next section. Both the Lagrangian code and the N-body code cover the entire time during which planetesimals are present beyond $r_{\rm{domain}}$.

\subsection{Last growth phase}\label{sec:last}
As discussed in \se{mercury}, bodies are removed from the N-body code when they migrate interior to $r_{\rm{domain}}$, to reduce computational time. After protoplanets have left the N-body code domain, information about their dynamics is lost. Therefore, we do not obtain the eventual planet period ratios, and we make the assumption that the planets leaving the N-body code domain keep their order and do not merge. However, we can still compute their final masses and water fractions, which is what we are ultimately after. In order to do this, we have constructed a semi-analytic model of the very last pebble accretion phase (`A-code'; \fg{completecartoon}). During this last growth phase, protoplanets are in the dry region. We assume that the planets are on perfectly circular and planar orbits because we do not have the dynamical information on the planets in the A-code. Assuming zero inclinations is justified because tidal damping timescales are very short \citep{2014MNRAS.443..568T,2019arXiv190210062L}. However, due to resonant forcing, the planetary embryos are expected to have orbits with eccentricities $e_{p}$ on the order of the disk aspect ratio $h$ \citep{2014AJ....147...32G,2014MNRAS.443..568T}. We check in \se{ecce} that our results do not change significantly if we assume $e_{p} = h$ instead of $e_{p} = 0$ in the pebble accretion efficiency calculations for the very inner disk.

Each time a protoplanet has crossed $r_{\rm{domain}}$, we integrate the mass growth rate of each planet interior to $r_{\rm{domain}}$, $d M_{\rm{planet}, i} / dt$, in time, until the next protoplanet crosses $r_{\rm{domain}}$ and is added to the sequence of protoplanets in the semi-analytic model. The growth rate of protoplanet $i$ is given by
\begin{equation}
\frac{d M_{\rm{planet}, i}}{dt} = \epsilon_{\rm{PA}, i} (t) \: \dot{M}_{\rm{peb, r_{\rm{domain}}}} (t) \: \prod_{j = 1}^{j = i - 1}{1 - \epsilon_{\rm{PA}, j} (t)},
\end{equation}
where $i \in [1, N]$ with $N$ the total number of protoplanets that crossed $r_{\rm{domain}}$ (an incremental function of time), and where $i = 1$ and $i = N$ correspond to the outermost and the innermost protoplanet in the A-code, respectively. The symbol $\dot{M}_{\rm{peb, r_{\rm{domain}}}}$ denotes the pebble mass flux at the inner boundary of the N-body code domain $r_{\rm{domain}}$. The value of $\dot{M}_{\rm{peb, r_{\rm{domain}}}}$ follows (as a function of time) from the N-code and the L-code. To calculate the pebble accretion efficiency $\epsilon_{\rm{PA}}$ in the A-code we use the same expression as used in the N-code, which is based on \citet{2018A&A...615A.138L} and \citet{2018A&A...615A.178O}. 

In calculating $\epsilon_{\rm{PA}}$, we assume that the protoplanets stall at $r_{\rm{domain}}$ \hbox{($\sim$0.07~au)}. We check, however, that putting the planets at random (but ordered) locations between $r_{\rm{domain}}$ and the current position of the innermost TRAPPIST-1 planet ($\sim$0.01~au) during the integration of the growth rates does not matter much for the final outcome, because the pebble accretion efficiency does not vary much between 0.01~au and 0.07~au (because the disk aspect ratio $h = H_{\rm{gas}} / r$ is nearly constant over this distance). Therefore, planetary migration in the final growth phase followed by the A-code is not so important for the final planet masses and compositions.

When a planet grows large enough to induce a pressure bump exterior to its orbit, the pebble flux is halted: no pebbles can reach the planet's orbit. The mass at which this occurs is called the pebble isolation mass \citep{2012A&A...546A..18M,2014A&A...572A..35L,2018A&A...612A..30B,2018A&A...615A.110A}.    
We do runs both including pebble isolation and without. When we include pebble isolation, the growth of a protoplanet stalls when that protoplanet reaches the pebble isolation mass, and no pebbles reach protoplanets interior to the pebble-isolated planet.

\section{Results}
\subsection{Fiducial simulation}\label{sec:results}
We have chosen the fiducial values of our model parameters equal to the values used in \citet{OLS2017}. Our aim is not to perfectly reproduce the \hbox{TRAPPIST-1} system, for multiple reasons. First of all, the exact compositions of the \hbox{TRAPPIST-1} planets are uncertain, and we do not account for the possibility of water loss. Due to water loss, simulated planetary systems that we would consider too wet to resemble the \hbox{TRAPPIST-1} system could become drier systems post formation (see \se{loss} for further discussion). Secondly, computational times are too long to run an optimisation model (e.g. a Markov chain Monte Carlo model), for which many realisations are necessary. Thirdly, it is not trivial to define a quantitative metric for the similarity between synthetic and observed planetary systems (but see \citet{2019arXiv190109719A} for a pioneering work). Fourthly, the stochastic component in our model, the injection location of planetesimals in the N-body code (\se{couple}), leads to quite some variations between results of simulations with the same initial conditions, as we will demonstrate shortly. Therefore, with `typical TRAPPIST-1 system' we generally mean a compact system of several (5--15) planets, which have moderate water fractions on the order of 10\%.

The input parameters of our model are the gas accretion rate $\dot{M}_{\rm{gas}}$, the total gas disk mass $\rm{M}_{\rm{disk}}$, the disk outer radius $r_{\rm{out}}$, the metallicity $Z$, and the dimensionless turbulence strength $\alpha$, four of which are independent. The combination of $\dot{M}_{\rm{gas}}$ and $\alpha$ defines the gas surface density profile (\eq{sigma}). The metallicity of TRAPPIST-1 has been measured to be close to solar \citep{2018MNRAS.475.3577D} and we therefore take $Z = 0.02$. Observed gas accretion rates of low-mass M-dwarf stars are typically on the order of $10^{-10}$ solar masses per year \citep{2017A&A...604A.127M}, which is our fiducial value. Observations of gas disk outer radii are notoriously difficult. Setting the total gas disk mass to $\rm{M}_{\rm{disk}} = 0.038 \: \rm{M}_{\rm{Trappist-1}}$ leads to an outer disk radius $r_{\rm{out}} = 200$ au. The total solids budget of our fiducial disk is $Z \rm{M}_{\rm{disk}} = 18.3 \: \rm{M}_{\oplus}$, half of which is rocky material. Our fiducial model parameter values are listed in \tb{inputpar}.

\begin{table}
\caption{Fiducial model parameters.}
\label{tab:inputpar}
\centering
\begin{tabular}{l l l}
\hline
Gas accretion rate & $\dot{M}_{\rm{gas}}$ &$10^{-10} \: \rm{M}_{\odot} \rm{yr}^{-1}$\\
Turbulence strength & $\alpha_{\rm{T}}$ &$1 \times 10^{-3}$\\
Total disk mass (gas) & $\rm{M}_{\rm{disk}}$ & $0.038 \rm{M}_{\rm{Trappist-1}}$\\
Metallicity & $Z$ &0.02\\ 
Total disk mass (solids) & $\rm{M}_{\rm{solids}}$ & $18.3 \rm{M}_{\oplus}$\\
Disk outer radius & $r_{\rm{out}}$ & 200 au\\
\hline
\noalign{\vskip 3mm}   
\multicolumn{3}{l}{\begin{small}{\bf Notes.} The disk outer radius $r_{\rm{out}}$ follows from the choices for $\dot{M}_{\rm{gas}}$,\end{small}}\\
\multicolumn{3}{l}{\begin{small}$\alpha$, and $\rm{M}_{\rm{disk}}$.\end{small}}\\
\end{tabular}
\end{table}

\subsubsection{First growth phase} \label{sec:wet}
In this section we present the results of our fiducial model for the first, predominantly wet growth phase, with which we mean the growth and migration of planetesimals that we follow with the Lagrangian dust evolution code and the N-body code (represented by the two top rows in \fg{coupling}). The results of the semi-analytic procedure for the final, dry growth phase after protoplanets have crossed the inner domain radius of the N-body code (\se{couple}) are discussed in \se{dry}.

\begin{figure}
        \centering
                \includegraphics[width=0.49\textwidth]{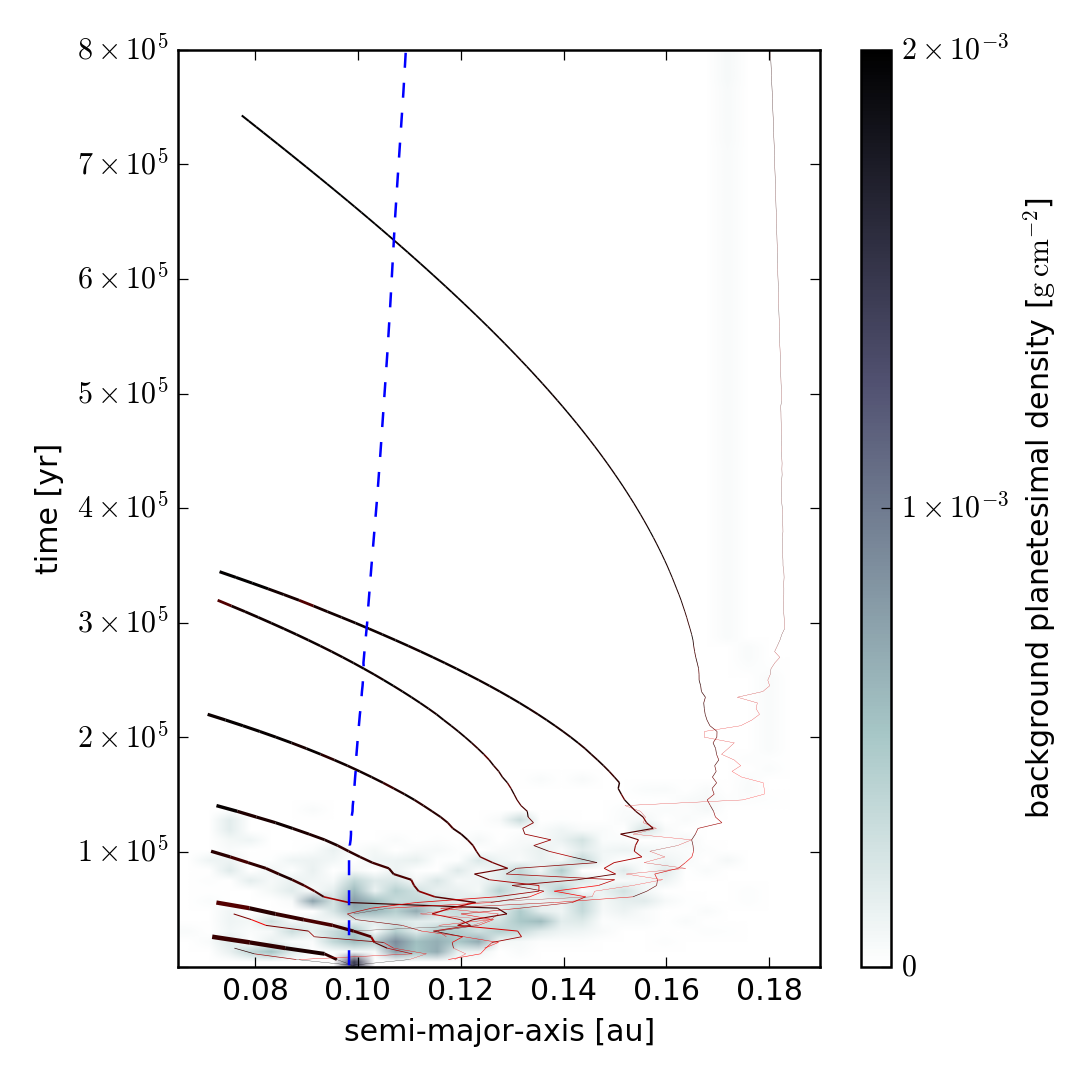}
\caption{Protoplanet trajectories (solid lines) for one realisation of the fiducial model. The water snowline (depicted by the dashed blue line) is located at around 0.1 au and moves outwards in time due to a decreasing icy pebble flux. Shaded regions depict background planetesimal densities (the protoplanets are not accounted for in these background planetesimal densities). The first planetesimals form after $\sim$$10^{3}$ years. The colour of the solid lines corresponds to the eccentricity of that protoplanet: the redder, the more eccentric its orbit; the blacker, the more circular. The line width is proportional to protoplanet mass.}\label{fig:trajectories_fid}
\end{figure}

In \fg{trajectories_fid} we show the evolution trajectories of growing planets for our fiducial simulation parameters. The shading corresponds to the density of planetesimals: the darker a region is shaded, the higher the density of planetesimals in that region. The snowline is depicted by the dashed blue line, and moves outwards in time because the flux of icy pebbles delivering water vapour to the inner disk decreases over time. As in \citet{2018A&A...620A.134S}, we find that planetesimals form outside the snowline at an early time (in the \hbox{first $\sim$$10^{5}$~years}). We note that \fg{trajectories_fid} gives the impression that planetesimals are present at $t$ = 0. This is because the first planetesimals form after $\sim$$10^{3}$~years, which is close to $t = 0$ on the timescale plotted. The planetesimal belt that is initially narrow, quickly broadens due to scattering.

We find that not only large protoplanets migrate across the N-body inner domain radius $r_{\rm{domain}}$; small planetesimals cross this radius as well due to scattering. The solid lines in \fg{trajectories_fid} correspond to the largest bodies that amount to more than 99\% of the total mass that crosses $r_{\rm{domain}}$. These are the bodies for which we follow their subsequent dry growth phase (\se{dry}). The smallest bodies -- corresponding to less than 1\% of the total mass that crossed $r_{\rm{domain}}$ -- are ignored in the semi-analytic model of the last growth phase (A-code). In \fg{removal_histogram}, a histogram shows the masses of bodies that crossed $r_{\rm{domain}}$, from a total of ten simulations with the fiducial input parameters. The red line denotes the cumulative mass percentage of bodies that crossed $r_{\rm{domain}}$. The vertical dotted line corresponds to the threshold below which bodies correspond to 1\% of the total mass that crossed $r_{\rm{domain}}$.

\begin{figure}
        \centering
                \includegraphics[width=0.49\textwidth]{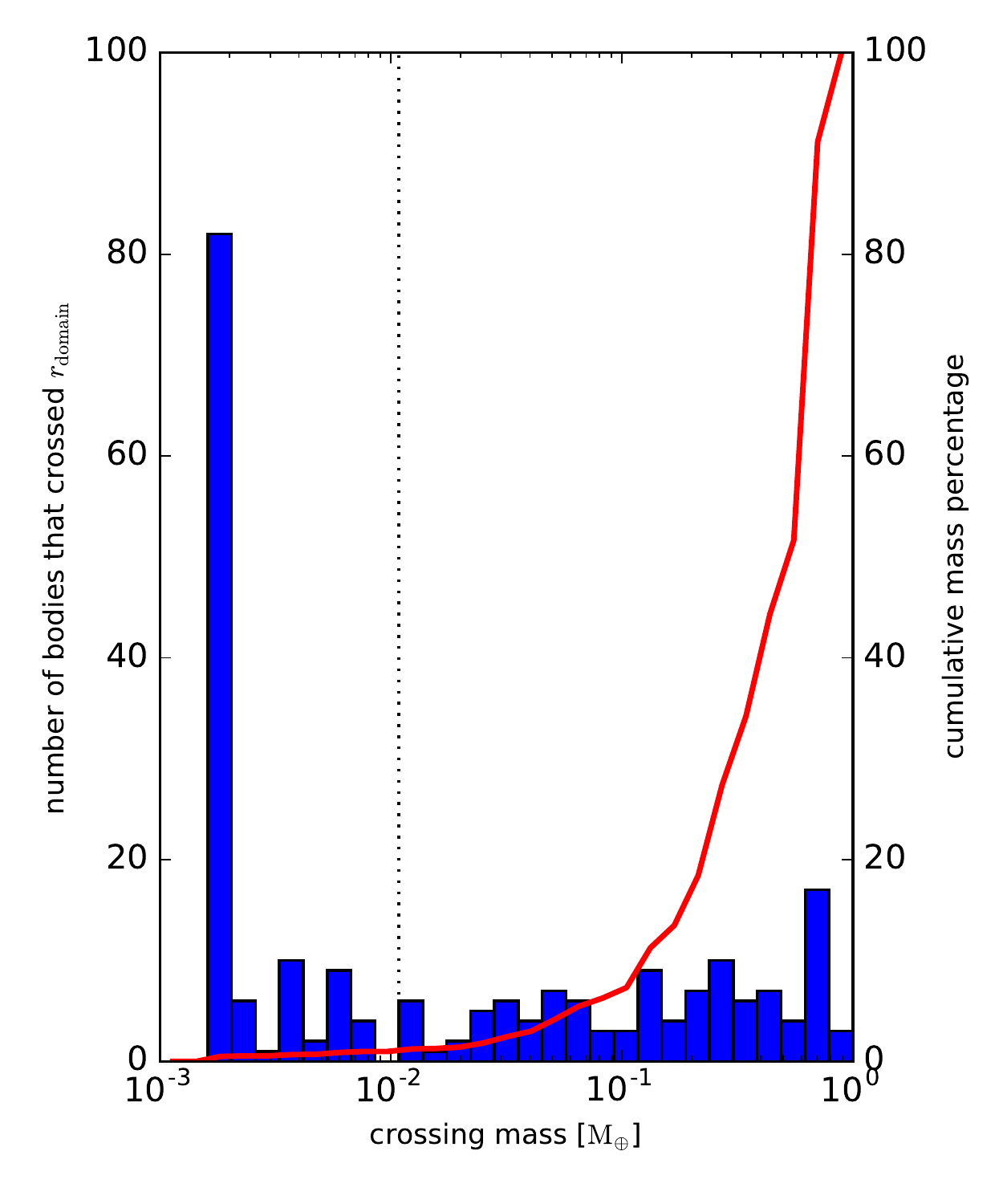}
\caption{Histogram of the masses of bodies that crossed the inner domain radius of the N-body simulation $r_{\rm{domain}}$. The peak at the smallest mass bin is due to bodies of the initial planetesimal size. The red line denotes the cumulative mass percentage of crossed bodies. The vertical dotted line corresponds to a cumulative mass percentage of 1\%. Bodies to the left of this threshold are neglected after they crossed $r_{\rm{domain}}$; only the bodies more massive than the threshold mass are followed. This plot stacks the results of ten realisations of the fiducial simulation.}\label{fig:removal_histogram}
\end{figure}

\begin{figure}
        \centering
                \includegraphics[width=0.49\textwidth]{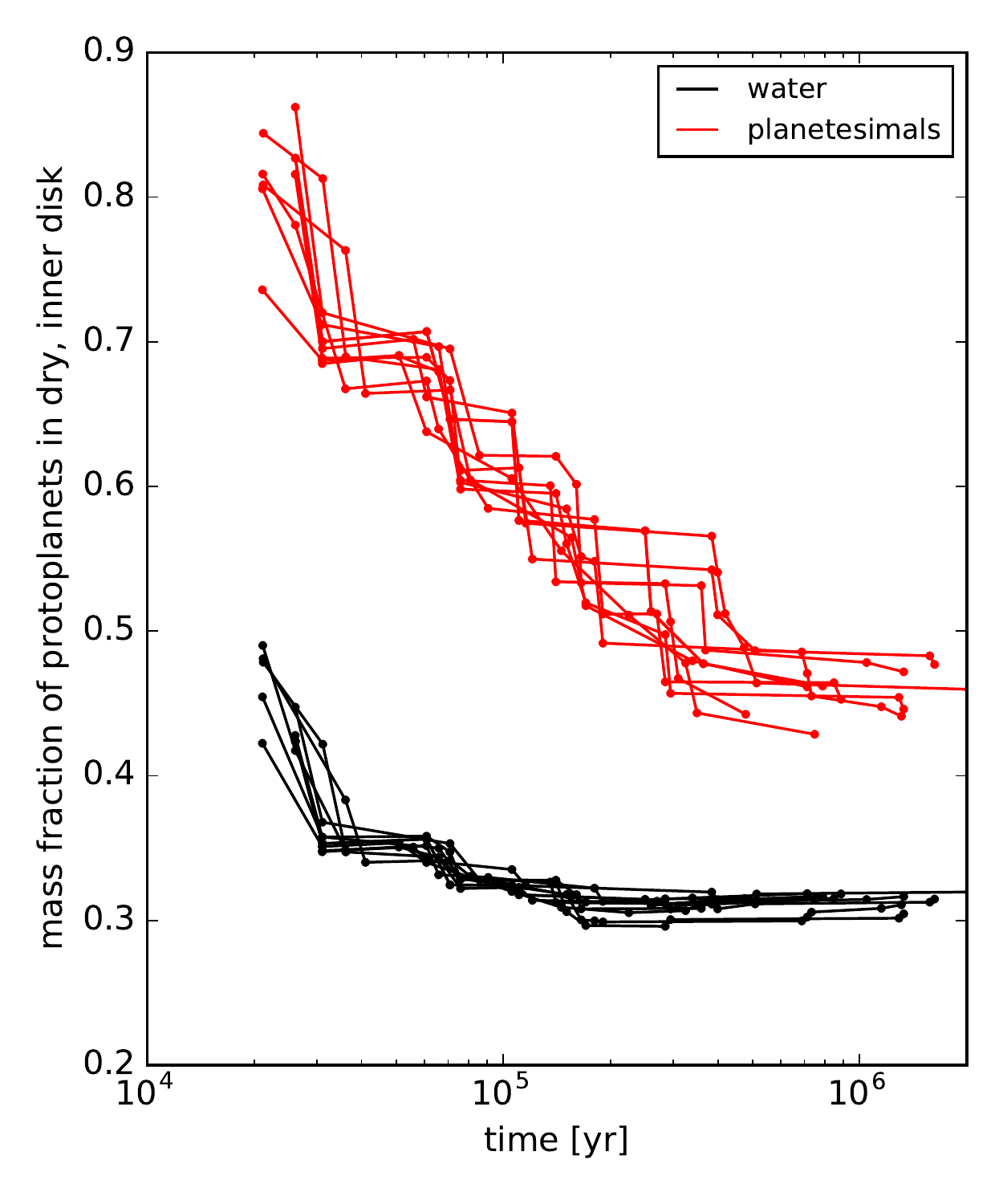}
\caption{Water mass fraction (black) and planetesimal mass fraction (red) of the total mass in protoplanets (and planetesimals) that have crossed the inner domain radius of the N-body simulation $r_{\rm{domain}}$, as a function of time. Here we have not yet taken into account the final growth stage in the dry inner disk (A-code). Results of ten simulations with the fiducial parameters are plotted. Each crossing event corresponds to a red and a black dot. Crossing events belonging to the same simulation are connected by a line.}\label{fig:mass_fractions}
\end{figure}

The masses of protoplanets are accreted from pebbles as well as from planetesimals. The planetesimal fraction (the mass fraction contributed by planetesimal accretion) and the water fraction of the total amount of bodies that crossed $r_{\rm{domain}}$ are plotted as a function of time in \fg{mass_fractions}. Each line corresponds to one realisation of the fiducial simulation, and each `crossing event' is denoted with a dot. We find that the general trend is that the planetesimal mass fraction of the mass that crossed $r_{\rm{domain}}$ (red lines) goes down with time. This is because the later bodies cross the domain radius $r_{\rm{domain}}$, the more their growth has been dominated by pebble accretion rather than planetesimal accretion. This is also reflected in \fg{trajectories_fid}, where it is clear that the bodies that migrate inwards at a relatively late time, have been scattered outwards at early times. After having been scattered outwards, these protoplanets find themselves in a planetesimal-free region, and grow mostly by pebble accretion. In contrast, the protoplanets that migrate inwards at an early time have grown in a planetesimal-rich region, and have therefore grown mostly by planetesimal accretion rather than pebble accretion. This is also the reason why in \fg{mass_fractions} the water fraction of the total mass in bodies that crossed $r_{\rm{domain}}$ (black lines) goes down with time. Whereas wet planetesimals were available for the `early migrators' even interior to the snowline, `late migrators' could only accrete dry pebbles after migrating past the snowline.

\subsubsection{Last growth phase}\label{sec:dry}
After protoplanets migrate across the N-body inner domain radius $r_{\rm{domain}}$, we select the largest bodies (corresponding to $>$99\% of the total mass in bodies that crossed $r_{\rm{domain}}$, see \fg{removal_histogram}) and calculate their final growth stage using a semi-analytic model (A-code; \se{last}). In this last growth phase, protoplanets are well interior to the snowline and grow by accreting dry pebbles.

\begin{figure*}
        \centering
                \includegraphics[width=0.49\textwidth]{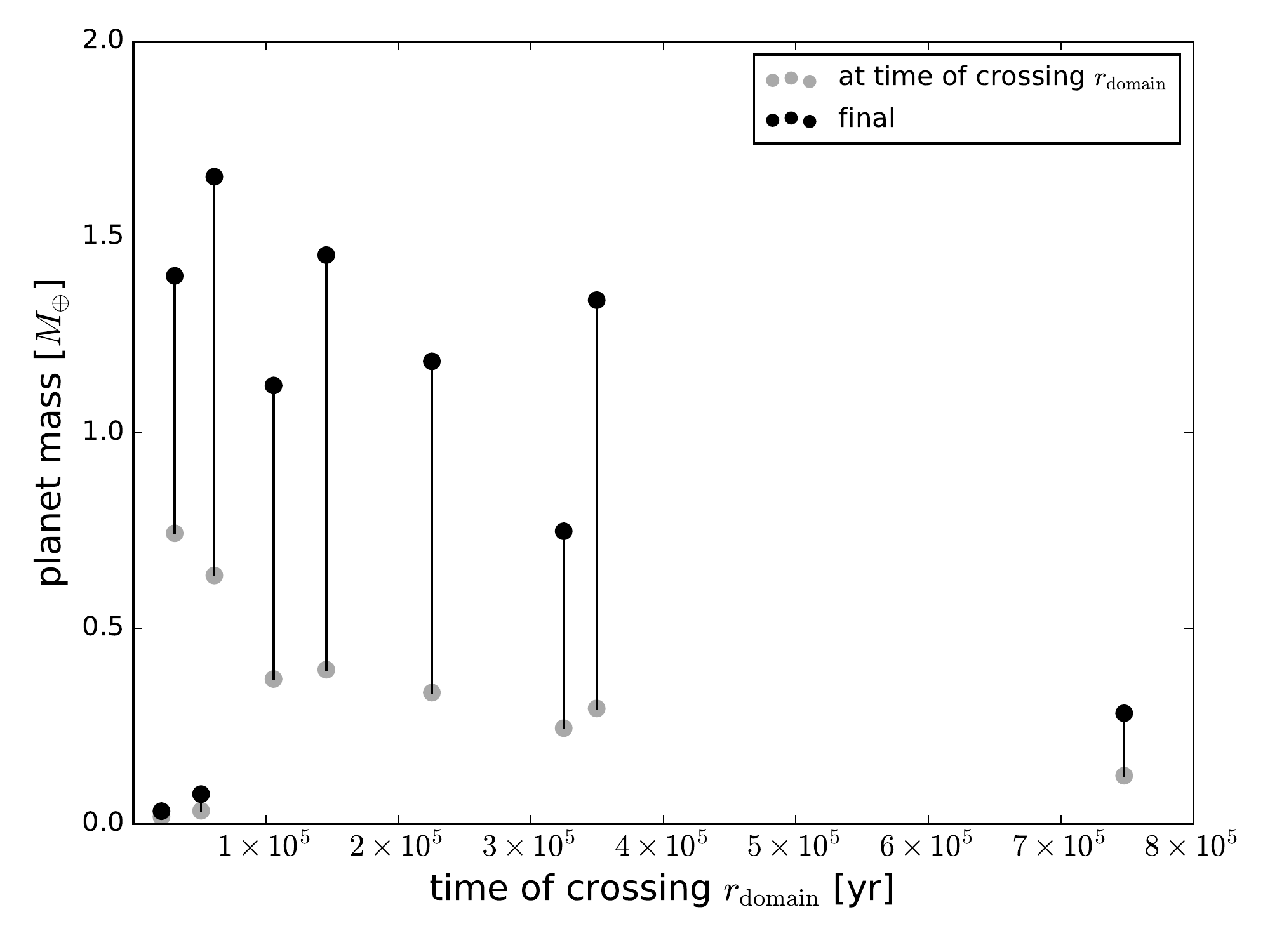}
                \includegraphics[width=0.49\textwidth]{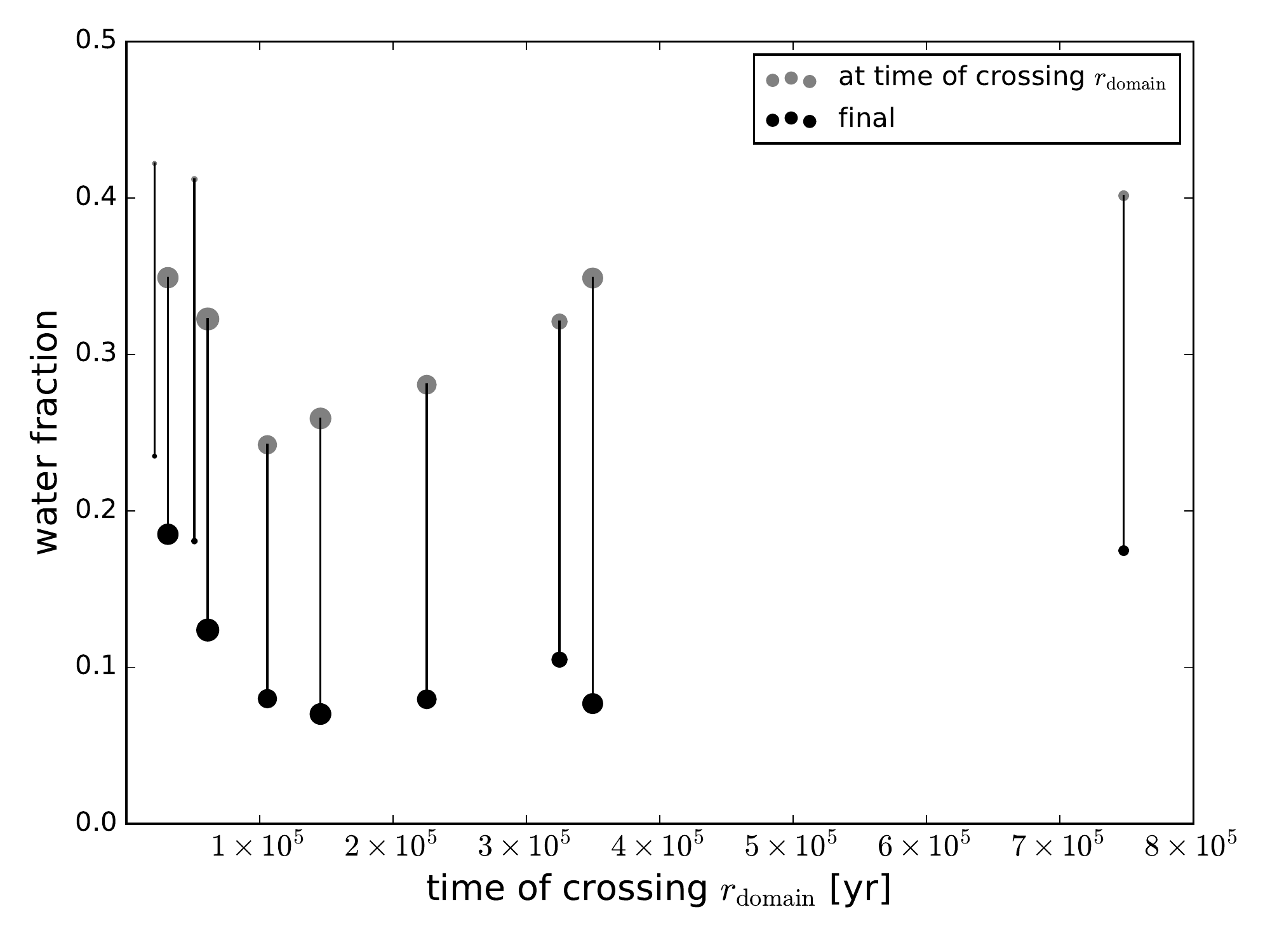}
\caption{Properties of simulated planets from one simulation with the fiducial model parameters. Left: Protoplanet masses at the time of their migration across the inner domain radius of the N-body code $r_{\rm{domain}}$ (grey) and their final masses (black), plotted against the time at which they migrated across $r_{\rm{domain}}$. Right: Water mass fractions of protoplanets at the time of their migration across the inner domain radius of the N-body code $r_{\rm{domain}}$ (grey) and their final water fractions (black), plotted against the time they crossed $r_{\rm{domain}}$. In the right panel, the dot size is proportional to the final planet mass.}\label{fig:drygrowth_fid}
\end{figure*}

In \fg{drygrowth_fid} we show the results of the semi-analytic computation of this last growth phase for one simulation with the fiducial model parameters. In this calculation we have not taken into account a pebble isolation mass, so for a given protoplanet, pebble accretion continues until the pebble mass flux at that protoplanet's position dries out.
In the left panel, the masses of protoplanets at the time of their crossing the N-body inner domain radius $r_{\rm{domain}}$ are plotted against their crossing time by the grey dots, and their final masses are plotted by the connected black dots. We note that by virtue of our model, the ordering in crossing times is the same as the final ordering in planets: the planet that migrated across $r_{\rm{domain}}$ first (last) ends up as the innermost (outermost) planet in the system. We see that the first and the third planet are much less massive than the other eight planets, with masses of 0.03 and 0.06 Earth masses, respectively. Their masses at $r_{\rm{domain}}$ just exceeded our threshold mass (\fg{removal_histogram}). Due to their relatively low masses, they are less efficient at accreting pebbles than the more massive protoplanets, and therefore their masses stay low. We also note that even though the second planet entered the last growth phase with a higher mass than the fourth planet, the fourth planet eventually becomes more massive than the second planet. This is because the pebble flux at the second planet's position dries out more quickly, due to exterior planets accreting pebbles, than the pebble flux at the fourth planet's position. 

The right panel of \fg{drygrowth_fid} shows the water fractions of the planets at the time at which they migrated across the inner domain radius of the N-body simulation $r_{\rm{domain}}$ (in grey) as well as their final water fractions (in black), plotted against the time they crossed $r_{\rm{domain}}$. In this plot the dot sizes are proportional to the final masses of the planets. The water fractions go down during the last growth phase because planets only accrete dry pebbles. The first and the third planet start and end the last, dry growth phase being more water-rich than the other eight planets. This reflects the fact that these planets crossed $r_{\rm{domain}}$ at relatively low masses, consisting predominantly of wet planetesimals. 

Ignoring the two smallest planets for now, we find that the innermost and outermost planets generally end up more water-rich than the middle planets, leading to a `V-shape' in the planet water fraction with orbital distance. With orbital distance (again ignoring the two smallest planets), the planet water fraction starts relatively high at $\sim$0.19 for planet c, gradually goes down until it reaches its lowest value of $\sim$0.07 at planet g, beyond which it goes up again to a value of $\sim$0.17 at planet k. Planet j is an exception to the V-shape, as we discuss below. In \fg{bars} we plot the constituents of each planet (wet planetesimals, wet pebbles, and dry pebbles) for each planet from this simulation. The `V-shape' is also visible in this figure. The planets that formed earliest (the innermost planets) accreted the most (wet) planetesimals (\se{wet}; \fg{mass_fractions}). The later a planet formed, the more its growth was dominated by pebble accretion. The middle planets are therefore generally the driest, because they accreted a lot of dry pebbles after crossing the snowline. The outermost planets also grew mostly by pebble accretion, but by the time they crossed the snowline and could accrete dry pebbles, the pebble flux had already dried out to a great extent. In this simulation, the second-most outer planet does not obey this trend of water fraction with planet order. Because the outermost planet formed relatively late, the second-outermost planet had quite a large pebble flux to accrete from for quite a long time, without an exterior planet `stealing' part of the pebbles.

\begin{figure}
        \centering
                \includegraphics[width=0.49\textwidth]{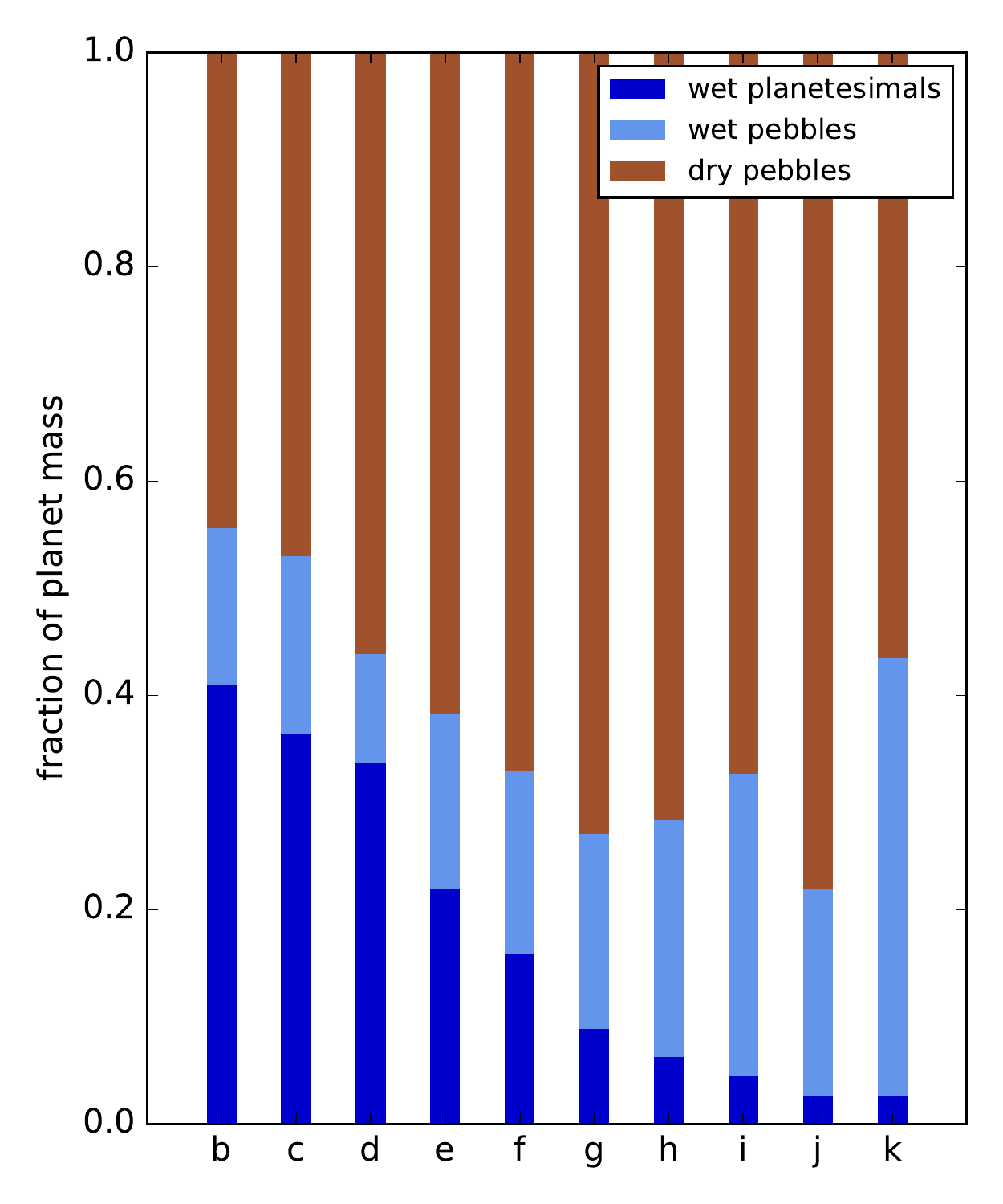}
\caption{Stacked bar chart of planet constituents for each planet of the fiducial model run. Dark blue bars correspond to planetesimals (which are formed outside the snowline and therefore contain $50\%$ water); light blue bars denote wet pebbles (also $50\%$ water) accreted outside the snowline; brown bars depict dry pebbles accreted interior to the snowline. The `V-shape' in the water fraction with planet order (right panel of \fg{drygrowth_fid}) is visible in this figure as well.}\label{fig:bars}
\end{figure}

\subsubsection{Comparison to the TRAPPIST-1 system}
In \fg{mass_radius} we compare the simulated planets from the fiducial simulation with the TRAPPIST-1 planets on a mass-radius diagram. The solid, dotted, and dashed lines correspond to the mass-radius relations for a rocky composition with 0\%, 10\%, and 20\% water mass fractions, respectively \citep{2018ApJ...865...20D}. The \hbox{TRAPPIST-1} planet data are taken from \citet{2018ApJ...865...20D}. The two smallest simulated ``planets'', the innermost (`b') and the third innermost (`d') (see \fg{drygrowth_fid}), fall outside the mass domain plotted here (it is conceivable that in reality these small bodies were swallowed by the other, bigger planets). The arrows starting from each simulated planet data point depict the trajectory that each simulated planet would follow on the mass-radius diagram when its water content evaporated completely (see \se{loss}). Earth and Venus are plotted by the black dots and fall on the mass-radius line corresponding to a water fraction of 0\%. The water fractions of simulated planets are similar to those of the \hbox{TRAPPIST-1} planets. The simulated planetary system contains four planets that are slightly more massive than the most massive TRAPPIST-1 planet, but the simulated system agrees with our rough definition of a `typical TRAPPIST-1 system', which is quite remarkable given our lack of parameter optimisation. 

\subsubsection{Multiple realisations}
In \fg{ninetimes} the mass-radius diagrams for nine other realisations of the fiducial simulation are shown. Different realisations of the same simulation do not lead to identical results. The scatter is due to the stochastic component of our model, which is the exact injection location of planetesimals in the N-body code. However, the characteristics of the resulting planetary systems are similar. All realisations of the fiducial simulation lead to a number of planets between 9 and 13 with masses between 0.02 and 3.6 Earth mass.
In \tb{results} we list the planetary system characteristics, such as the number of planets and the mean planet water fraction, whic result from different model variations. For each quantity, the mean and the standard deviation are provided, based on the results of different realisations of each model. The mean and standard deviation of the individual planet mass and water fraction are weighted by the individual planet mass, in order to reduce the importance of very small planets to these quantities (which in reality may not have survived anyway). The results of our fiducial simulation are listed in the first row. The second row provides results for the fiducial model where we have accounted for the pebble isolation mass, as we discuss in \se{pin}. In \se{parameters} we discuss the results of the other model runs, in which we have varied several input parameters.

\begin{figure*}
        \centering
                \includegraphics[width=0.98\textwidth]{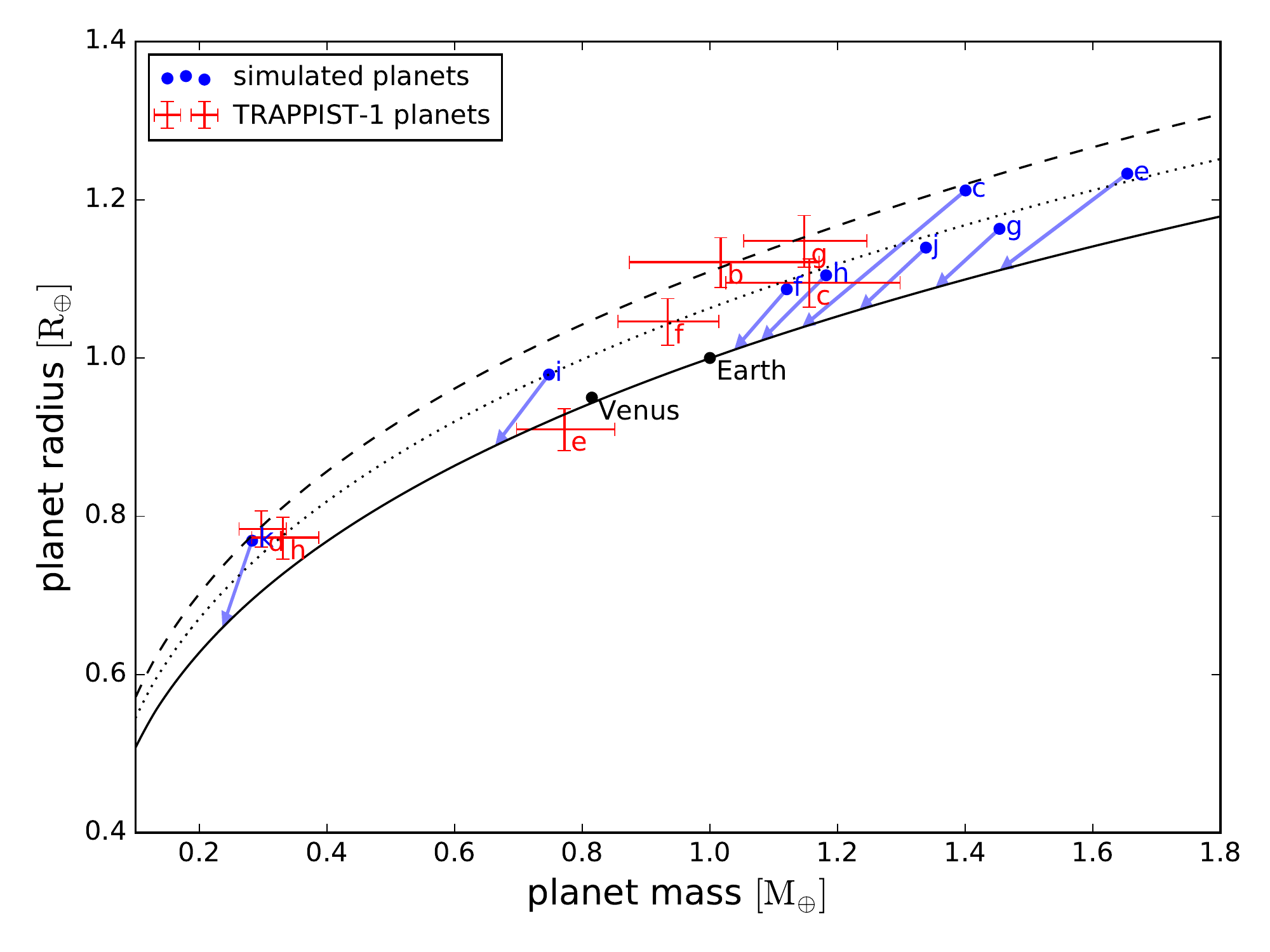}
\caption{Mass-radius diagram of simulated planets (blue dots) and TRAPPIST-1 planets (red crosses), data taken from \citet{2018ApJ...865...20D}. The solid, dotted, and dashed lines correspond to mass-radius relations for a rocky composition and a water fraction of 0\%, 10\%, and 20\%, respectively. The mass-radius relations are the same as the ones used in \citet{2018ApJ...865...20D}. The arrows starting from the simulated planets depict the trajectories on the mass-radius diagram that the simulated planets would cover if they were to lose their entire water content.}\label{fig:mass_radius}
\end{figure*}

\subsubsection{Pebble isolation mass}\label{sec:pin}
Our fiducial simulations have produced planets with masses of approximately Earth mass without the implementation of the pebble isolation mass (PIM). Taking the PIM into account imposes a maximum mass on the planets. The PIM has been parametrised as a function of the disk aspect ratio $h = H_{\rm{gas}} / r$, the turbulence strength $\alpha$, and the stopping time of pebbles $\tau$ by using hydrodynamical simulations (e.g. \citet{2014A&A...572A..35L,2018A&A...612A..30B, 2018A&A...615A.110A}). Filling in the expressions for the PIM provided by \citet{2018A&A...612A..30B} and \citet{2018A&A...615A.110A}, we find PIM values of 5.0 and 1.8 Earth masses at 0.05 au for our fiducial disk model, respectively. Because these values differ by a factor of $\sim$2, and are larger than the \hbox{TRAPPIST-1} planet masses, we here assume a pebble isolation mass of one Earth mass, independent of location\footnote{An additional complication in using a physical expression for the PIM would be the dependency on the disk aspect ratio, which in our model varies slightly with semi-major axis, and we do not know the exact orbital distances of the protoplanets in the A-code.} and time, with the goal of demonstrating the effect of imposing a maximum planet mass. We take our fiducial simulation results and run the semi-analytical model (A-code), in which we now limit the planet growth to \hbox{1 $M_{\oplus}$}. 

In the left panel of \fg{pin}, the water fractions of the planets at the time at which they migrated across $r_{\rm{domain}}$ (in grey) as well as the resulting final water fractions (in black), are plotted against their crossing times. The dot sizes are proportional to the final planet masses. We find that compared to the fiducial simulation without a PIM (\fg{drygrowth_fid}), the nine innermost planets are more water-rich. Five of these nine planets have reached the PIM, so that they have accreted fewer dry pebbles than in the case without a PIM, leading to higher final water fractions. The other four planets have been starved of pebbles once an exterior planet became pebble isolated, and have thus also accreted fewer dry pebbles than in the no-PIM case. The outermost planet has not reached PIM nor has it been starved of pebbles by an exterior pebble-isolated planet, and therefore it has the same final water fraction as in \fg{drygrowth_fid}. In the right panel of \fg{pin} the mass-radius diagram is presented for the simulation including the PIM. The simulated planets that were too massive to match the TRAPPIST-1 planets in our fiducial model without a pebble isolation mass (\fg{mass_radius}) are now limited to one Earth mass. In \tb{results} the general results for our simulations including the PIM are listed (`model 2').

\begin{figure*}
        \centering
                \includegraphics[width=0.49\textwidth]{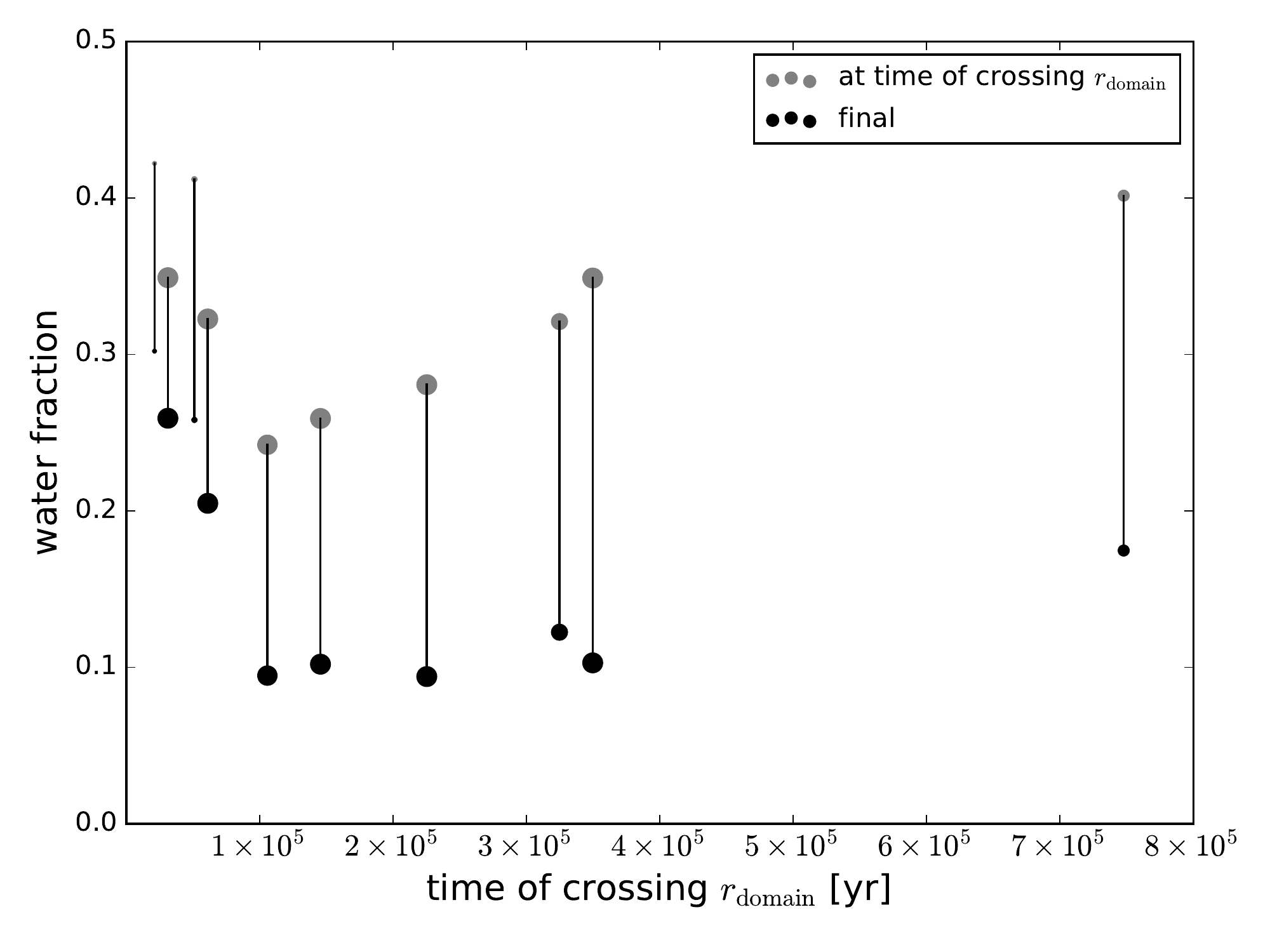}
                \includegraphics[width=0.49\textwidth]{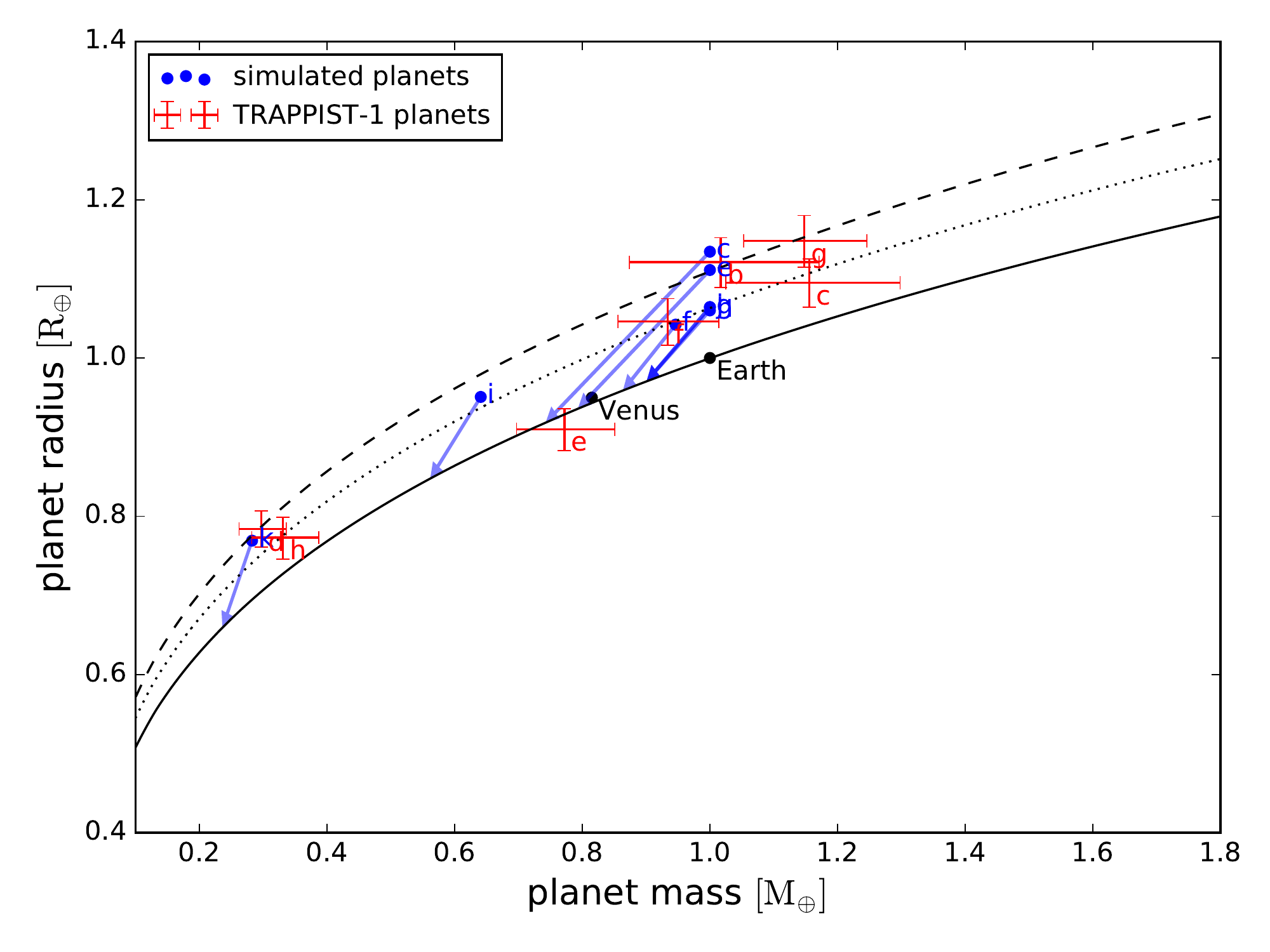}
\caption{Properties of planets from the fiducial simulation including a pebble isolation mass of 1~$M_{\oplus}$. Left: Same as the right panel of \fg{drygrowth_fid}, but now including a pebble isolation mass of 1 $M_{\oplus}$. Right: Same as \fg{mass_radius}, but now including a pebble isolation mass of 1 $M_{\oplus}$.}\label{fig:pin}
\end{figure*}

\subsubsection{Non-zero eccentricity in very inner disk}\label{sec:ecce}
Due to orbital resonances, the protoplanets in the very inner disk are expected to have non-zero eccentricities, on the order of the disk aspect ratio: $e_p \sim h$ \citep{2014AJ....147...32G,2014MNRAS.443..568T}. We have already discussed that the exact orbital distance of a protoplanet in the very inner disk ($\sim$$0.01$--$0.07$ au, covered by the A-code) does not matter much for the pebble accretion efficiency, as the disk aspect ratio $h$ is nearly constant over this distance range. Therefore, as long as the ordering of the chain of planetary embryos remains intact, the exact locations of the embryos in the very inner disk are not important for the final planet masses and water fractions. We now check whether our results change when we take $e_p = h$ into the calculations of the pebble accretion efficiencies in the very inner disk, instead of assuming zero eccentricities in the very inner disk (which is what we have done up to now). In \fg{ecce} we show that the results change only slightly if we take $e_p = h$, thereby justifying our assumption that planets in the very inner disk move on circular orbits when calculating their pebble accretion efficiencies. The reason why the results with $e_{p} = h$ are very similar to those with $e_{p} = 0$ is that the relative velocity due to eccentricity ($e_{p} v_{K})$ is not dominant over the Keplerian shear velocity ($\sim$0.01 $v_{K}$) between the planetary embryos and the pebbles \citep{2018A&A...615A.138L}. The pebble accretion efficiencies for $e_{p} > 0$ are always larger than for $e_{p} = 0$ because we kept the planet inclinations equal to zero \citep{2018A&A...615A.138L}. 

\begin{figure*}
        \centering
                \includegraphics[width=0.49\textwidth]{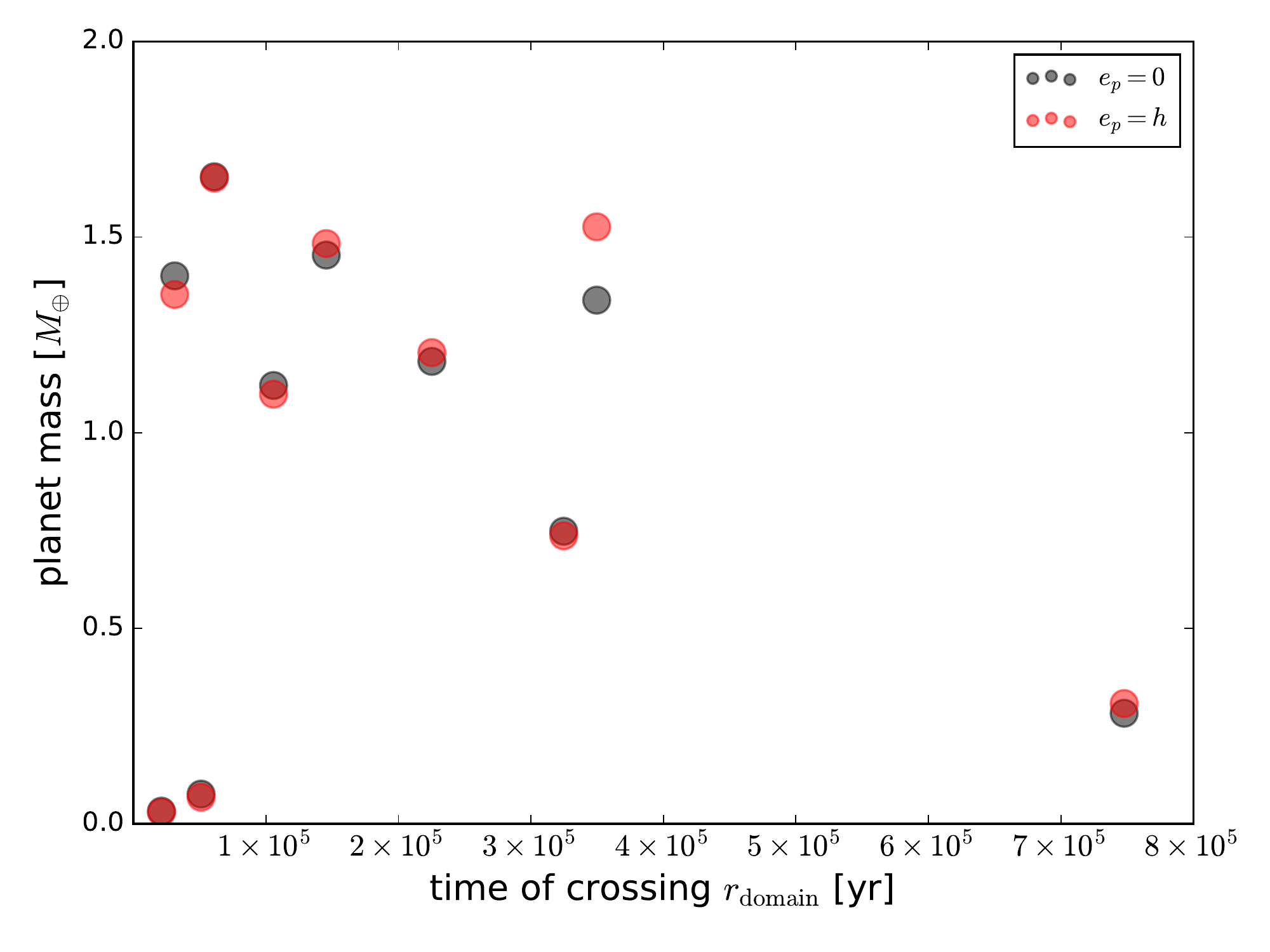}
                \includegraphics[width=0.49\textwidth]{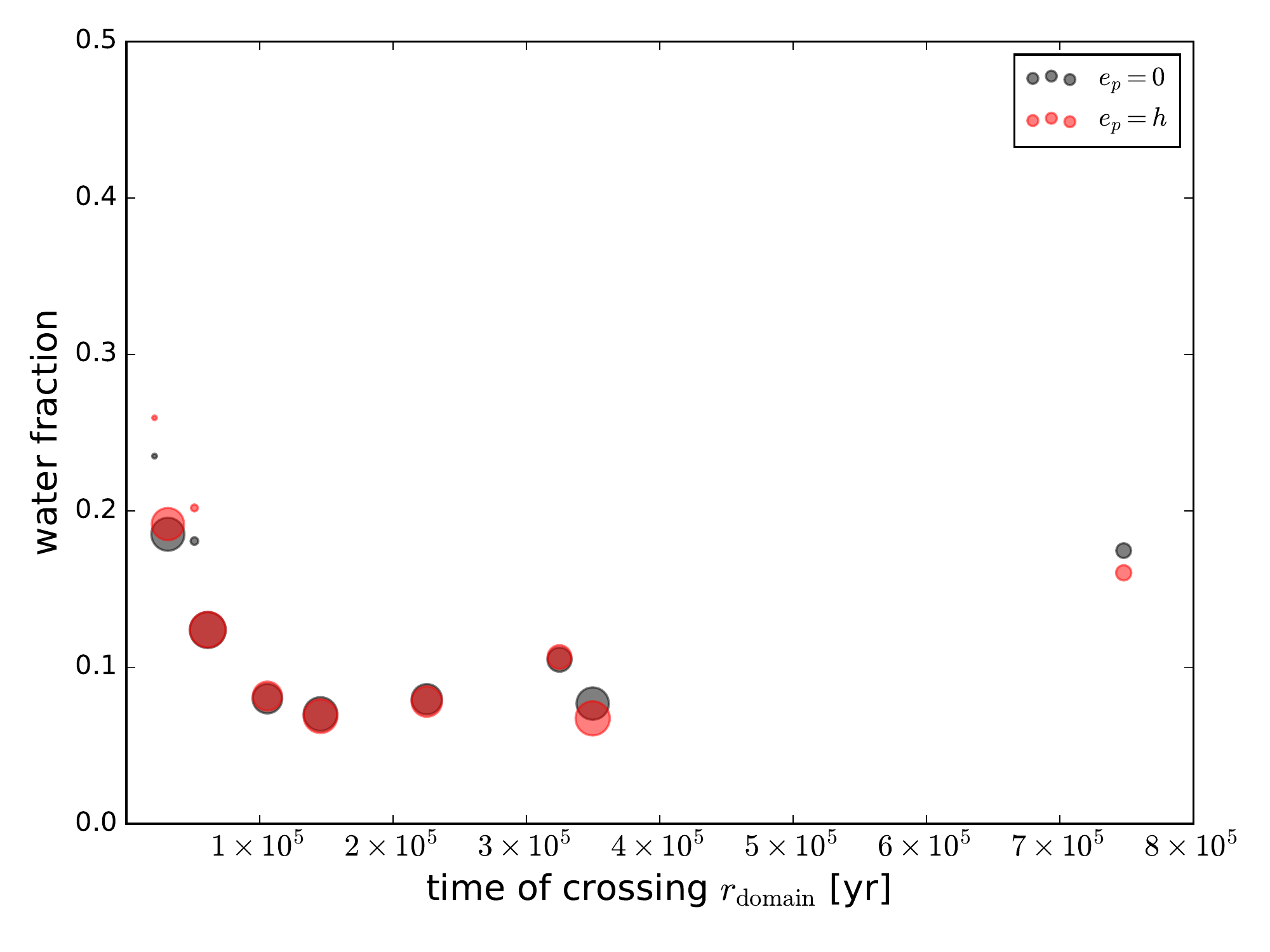}
\caption{Comparison of planet properties from the fiducial simulation and the fiducial simulation with non-zero protoplanet eccentricities. Left: Final planet masses against the time at which they migrated across $r_{\rm{domain}}$ for the fiducial simulation (black), and for the fiducial simulation accounting for non-zero protoplanet eccentricities $e_p$ in the very inner disk (by setting $e_p$ equal to the disk aspect ratio $h$ in the A-code) (red). Right: Same as left panel, but now for final planet water fractions instead of masses. Taking into account non-zero eccentricities in the very inner disk leads to only small variations in the final results.}\label{fig:ecce}
\end{figure*}

\subsection{Dependence on model parameters}\label{sec:parameters}

In this section we discuss the outcomes of simulations in which we vary one input parameter at a time. The results are listed in \tb{results}. In all models except `model 9' and `model 10', we keep the initial gas and solids disk masses fixed at the fiducial values (\tb{inputpar}), by adjusting the disk size accordingly.

We first check the dependence of our results on the initial planetesimal size, the fiducial value of which is 1200 km. Decreasing the initial planetesimal size to 1000 km (`model 4') does not change the results, which justifies our fiducial choice of 1200 km. Increasing the initial size to 1500 km (`model 3') leads to slightly fewer and larger planets, but the total mass in planets remains unchanged.

A lower value of the turbulence strength $\alpha$ implies a smaller disk, because the gas surface density $\Sigma_{\rm{gas}}$ is inversely proportional to $\alpha$ in our framework of a viscous gas disk. A lower \hbox{$\alpha$-value} also implies a larger pebble flux, because the dust surface density is initially a constant fraction of the gas surface density. Because we only vary $\alpha$ and keep the gas accretion rate $\dot{M}_{\rm{gas}}$ fixed, the pebble-to-gas flux ratio is also larger for smaller $\alpha$. Planetesimal formation by streaming instability outside the snowline is strongly dependent on the pebble-to-gas mass flux ratio \citep{SO2017}. It is much less dependent on the disk size, because solid material from the very outer parts of the disk does not take part in the streaming instability outside the snowline, which happens at an early stage of the evolution of the disk. Therefore, as was already noted in \citet{2018A&A...620A.134S}, even though we keep the solids budget fixed, a smaller $\alpha$ leads to the formation of more planetesimals (disregarding secondary effects such as the subtle dependencies on $\alpha$ of the pebble size and of the streaming instability threshold on the pebble flux). Indeed, decreasing the value of $\alpha$ to $5 \times 10^{-4}$ (`model 5') results in more mass in formed planetesimals $\rm{M}_{\rm{SI}}$. We note that although we lowered $\alpha$ by a factor of 2, the mass in formed planetesimals increased by a larger factor. This is because the planetesimal formation rate is proportional to the pebble flux $\dot{M}_{\rm{peb}}$ that is in excess of the critical pebble mass flux $\dot{M}_{\rm{peb,crit}}$ (\eq{excess}). Increasing the pebble flux by a factor of $f$ therefore results in the planetesimal formation rate being increased by a factor greater than $f$ \citep{2018A&A...620A.134S}. In the case of $\alpha = 5 \times 10^{-4}$, most of the icy pebble flux is converted to planetesimals. Due to computational limitations, we performed the simulations of this model variation with the larger initial planetesimal size of 1500 km, in order to curb the number of planetesimals in the N-body code. To get rid of the planetesimal-sized bodies that were scattered interior to the N-body domain, before focusing on the protoplanets in the semi-analytical calculation of the last growth phase (A-code), we had to neglect the smallest 3\%\ instead of 1\%.
On average, 45.8 planets form in these simulations, with an average total mass in planets of \hbox{14.8 $\rm{M}_{\oplus}$}. This is a lot more than in the fiducial model runs, because in this case there are many more planetesimal seeds to grow planets from, and pebble accretion is more efficient for lower $\alpha$.
The type-I migration timescale is inversely proportional to the gas surface density \citep{2002ApJ...565.1257T}, and is therefore shorter for lower $\alpha$, leading to faster migration across the snowline. Additionally, planetesimals form closer to the snowline for lower $\alpha$ \citep{SO2017}. From these considerations one would expect lower final water fractions for lower $\alpha$, because protoplanets outside the snowline have to traverse a shorter distance at a larger migration velocity to reach the inner disk in which they can accrete dry pebbles. However, the effect of the formation of many more wet planetesimals that accrete each other is more dominant, and we find that the planets resulting from the simulations with lower $\alpha$ have much higher water fractions and are smaller than those produced in the fiducial model\footnote{We note that we have not taken the pebble isolation mass into account here. The pebble isolation mass is lower for lower $\alpha$, which, depending on its exact value, might make the effect of having smaller planets with larger water fractions for lower $\alpha$ even stronger, by preventing protoplanets in the inner, dry disk from accreting dry pebbles.}.

\begin{table*}
\caption{Simulation results for different models.}
\label{tab:results}
\centering
\begin{tabular}{l l l l l l l}
\hline
\noalign{\vskip 0.2mm}   
& Model description & $\#$ planets & $\sum{\rm{M}_{\rm{pl}}} \: [\rm{M}_{\oplus}]$ & $\rm{M}_{\rm{SI}} \: [\rm{M}_{\oplus}]$ & $\overline{\rm{M}_{\rm{pl}}} \: [\rm{M}_{\oplus}]$ & $\overline{f_{\rm{H_{2}O}}}$\\
\noalign{\vskip 0.2mm}   
\hline
\noalign{\vskip 0.2mm}   
1. & Fiducial (see \tb{inputpar}) & $11.0 \pm 1.1$ & $9.3 \pm 0.08$ & $1.4 \pm 0.02$ & $1.74 \pm 0.89$ & $0.10 \pm 0.05$\\
2. & With a pebble isolation mass of 1 $\rm{M}_{\oplus}$ & $11.0 \pm 1.1$ & $5.5 \pm 0.56$ & $1.4 \pm 0.02$ & $0.84 \pm 0.29$ & $0.19 \pm 0.08$ \\
3. & Larger initial pltsml size (1500 km) & $10.8 \pm 1.7$ & $9.3 \pm 0.2$ & $1.4 \pm 0.04$ & $1.84 \pm 0.92$ & $0.10 \pm 0.05$\\
4. & Smaller initial pltsml size (1000 km) & $11.0 \pm 0.6$ & $9.4 \pm 0.08$ & $1.4 \pm 0.03$ & $1.74 \pm 0.73$ & $0.10 \pm 0.05$\\
5. & $\alpha = 5 \times 10^{-4}$ & $45.8 \pm 2.5$ & $14.8 \pm 0.06$ & $5.9 \pm 0.01$ & $0.67 \pm 0.28$ & $0.39 \pm 0.09$\\
6. & $\alpha = 2 \times 10^{-3}$ & $10.8 \pm 1.2$ & $8.7 \pm 0.17$ & $1.5 \pm 0.01$ & $1.43 \pm 0.71$ & $0.14 \pm 0.05$\\ 
7. & $\dot{M}_{\rm{gas}} = 5 \times 10^{-11} \: \rm{M}_{\odot} \: \rm{yr}^{-1}$ & $20.3 \pm 3.1$ & $10.8 \pm 0.46$ & $3.0 \pm 0.01$ & $1.06 \pm 0.50$ & $0.16 \pm 0.09$ \\
8. & $\dot{M}_{\rm{gas}} = 2 \times 10^{-10} \: \rm{M}_{\odot} \: \rm{yr}^{-1}$ & $14.6 \pm 1.2$ & $7.1 \pm 0.1$ & $1.4 \pm 0.04$ & $1.57 \pm 0.93$ & $0.13 \pm 0.06$ \\
9. & Higher disk mass ($r_{\rm{out}} = 300$ au) & $11.6 \pm 1.0$ & $14.5 \pm 0.49$ & $1.4 \pm 0.03$ & $2.56 \pm 1.53$ & $0.07 \pm 0.05$ \\
10. & Lower disk mass ($r_{\rm{out}} = 100$ au) & $10.0 \pm 0.6$ & $4.3 \pm 0.04$ & $1.3 \pm 0.02$ & $1.10 \pm 0.49$ & $0.17 \pm 0.05$\\
\hline
\noalign{\vskip 0.9mm}   
TRAPPIST-1 & UCM model, \citet{2018ApJ...865...20D} & 7 & $5.66^{+0.65}_{-0.61}$ &  & $0.95 \pm 0.26$ & $0.10 \pm 0.05$\\
\noalign{\vskip 0.9mm}   
\hline
\noalign{\vskip 3mm}   
\multicolumn{7}{l}{\begin{small}{\bf Notes.} The columns denote (from left to right): model description; total number of planets; total mass in planets; total mass in\end{small}}\\
\multicolumn{7}{l}{\begin{small}planetesimals formed by streaming instability; weighted mean of individual planet mass; mean planet water fraction (weighted\end{small}}\\
\multicolumn{7}{l}{\begin{small}with planet mass). Each column entry gives the mean and standard deviation of the corresponding quantity, calculated from\end{small}}\\
\multicolumn{7}{l}{\begin{small}multiple realisations. The model descriptions are elaborated on in the text.\end{small}}\\
\end{tabular}
\end{table*}

A higher value of $\alpha$, on the other hand, implies a larger disk and a smaller pebble flux that is sustained for a longer time. Because planetesimal formation happens only in the early stages of the disk evolution, one would expect fewer planetesimals to form in this case, due to the smaller pebble flux. However, increasing the value of $\alpha$ to $2 \times 10^{-3}$ (`model 6') results in slightly more planetesimals compared to the fiducial model. This is because although the pebble flux is smaller, the streaming instability phase lasts longer, due to the subtle dependency of the critical pebble mass flux (needed to trigger the streaming instability) on the balance between diffusion, vertical settling, and radial drift, and therefore on the combination of $\alpha$ and the dimensionless stopping time of pebbles $\tau$ \citep{SO2017, 2018A&A...620A.134S}. The migration timescale is longer for higher $\alpha$, and planetesimals are formed further away from the snowline. Therefore, protoplanets reside exterior to the snowline for a longer time compared to the fiducial model. These effects result in slightly higher final water fractions compared with the fiducial model. The final planet masses are a bit lower than in the fiducial model, because pebble accretion is less efficient for higher $\alpha$. For $\alpha$ values higher than $5 \times 10^{-3}$, the region outside the snowline does not reach the conditions for streaming instability and no planets are formed, but this could be resolved by changing other model parameters, such as the metallicity, as well.

Varying the gas accretion rate $\dot{M}_{\rm{gas}}$ has similar effects as varying $\alpha$. Lower (higher) $\dot{M}_{\rm{gas}}$ means a larger (smaller) disk with lower (higher) surface density and pebble flux. However, when we change $\dot{M}_{\rm{gas}}$, we also change the viscous temperature profile (\eq{vis}), such that the snowline is located further outwards (inwards) for higher (lower) $\dot{M}_{\rm{gas}}$. The pebble-to-gas mass flux ratio is inversely related to the snowline location \citep{2018A&A...620A.134S}. The planetesimal formation rate in turn depends on the pebble-to-gas mass flux ratio, but not linearly: we remember that only the pebble flux in excess of the critical pebble flux is converted to planetesimals (\eq{excess}). With all these complications, we find that reducing the gas accretion rate to $\dot{M}_{\rm{gas}} = 5 \times 10^{-11} \: \rm{M}_{\odot} \: \rm{yr}^{-1}$ results in approximately twice as much mass in planetesimals formed by streaming instability. As in the case of lower $\alpha$, this leads to a larger number of planets than in the fiducial case, because there are more planetesimal seeds from which to grow. The larger amount of wet planetesimals overshadows the effect of lower migration velocity (which tends to increase the planet water fraction), and the final water fractions are a bit higher than in the fiducial model. A higher gas accretion rate of $\dot{M}_{\rm{gas}} = 2 \times 10^{-10} \: \rm{M}_{\odot} \: \rm{yr}^{-1}$ in turn leads to an amount \hbox{of formed} planetesimals similar to the fiducial case. Because in this model the type-I migration timescale is shorter than in the fiducial model, the planets resulting from this model are slightly more numerous, less massive, and more water-rich than in the fiducial case.

In our model, the parameter values are fixed for the duration of the simulation. If in reality protoplanetary disks feature decreasing values of $\alpha$ during their evolution, we expect that planetesimal formation outside the snowline starts when the conditions for streaming instability are marginally reached, such that only a small fraction of the pebble flux is converted to planetesimals. Therefore, our fiducial model may be more physical than the models with lower and higher $\alpha$ values, in which larger fractions of the icy pebble flux were converted to planetesimals and planets formed with higher water fractions.

In `model 9' and `model 10' we change the disk outer radius $r_{\rm{out}}$ to 300 au and 100 au, respectively. The gas disk mass changes accordingly, in contrast to all models discussed up to now, in which we kept the gas disk mass fixed. Planetesimal formation takes place in the early stage of the disk evolution and is therefore not affected by these changes in the disk outer radius: the pebbles from which planetesimals are formed originate at distances to the star closer than 100 au. However, pebble accretion can go on for a longer (shorter) time if the disk is larger (smaller). This leads to more massive (less massive) planets for larger (smaller) $r_{\rm{out}}$ compared with the fiducial model. The water fractions for smaller $r_{\rm{out}}$ are also higher than in the fiducial model, because of less dry pebble accretion in the inner disk. We note that the opposite is not true: the mean water fraction for larger $r_{\rm{out}}$ is only slightly lower than that in the fiducial model. This is because a few planets `steal' the long-lasting flux of dry pebbles and become relatively water-poor, leaving the other planets without many dry pebbles to accrete; these other planets therefore stay water-rich.
We note that a model with a smaller disk size may be more realistic than our fiducial model, as observations seem to suggest that disks around very low-mass stars are quite small (<100 au) \citep{2013ApJ...764L..27R,2017ApJ...841..116H}, however, primordial disk radii may have been larger (e.g. \citet{2019MNRAS.486.4829R}).

Although the number of planets and their final masses and water fractions depend on our model parameters, all planetary systems resulting from our different model variations feature the same trend in the planet water fraction with planet order: the inner- and outermost planets are generally more water-rich than the middle planets, for the reasons explained in \se{dry}. For a very large disk (as in `model 9'), the uprise in water fraction in the outermost planets is less strong, because in that case dry pebble accretion is still important even for the outermost planets. 

\section{Discussion}\label{sec:discussion}
\subsection{Evolution of the snowline location}
In our model, the water snowline location depends on the temperature and vapour pressure. The vapour pressure depends on the influx of icy pebbles delivering water vapour to the inner disk, and therefore varies with time. The pebble flux goes down with time, and the snowline therefore moves outwards with time (\fg{trajectories_fid}). We have assumed that the gas accretion rate $\dot{M}_{\rm{gas}}$ and therefore the viscous temperature profile are constant in time. However, taking into account that the gas accretion rate decreases with time leads to an inward movement of the snowline \citep{2007ApJ...654..606G,2011ApJ...738..141O,2012MNRAS.425L...6M}. \citet{2018A&A...620A.134S} have shown that the planetesimal formation phase is not affected by the migration of the snowline, because the viscous evolution timescale is much longer than the duration of the planetesimal formation phase. The general picture of planets acquiring moderate water fractions by originating and growing outside the snowline and then migrating inwards would also not be affected by a moving snowline, as long as the early growth and migration of planets is faster. Starting with a single embryo of 0.01 Earth masses and a simplified planetary growth function, \citet{2019A&A...624A.109B} calculated planetary growth tracks and resulting planetary compositions in an evolving disk, in which the snowline moves inwards with time. When they only include inward migration, they find that embryos indeed grow and migrate inwards faster than the evolution of the snowline. However, as they mention, the eventual composition of a planet does depend on the growth rate (pebble flux). For example, one can imagine that late-stage pebble accretion might be affected by snowline migration, if the pebble flux coming from the outer disk remains considerable on the viscous evolution timescale. The outermost planets -- that migrate inwards at late times, when the snowline might have moved inwards considerably -- could then be surrounded by icy pebbles rather than dry pebbles for a slightly longer time. This would lead to even more water-rich outer planets and therefore to a strengthening of the trend in the water fraction with planet ordering (inner- and outermost planets being more water-rich than middle planets; see \se{dry}).

\subsection{Water loss}\label{sec:loss}
In our simulations we have not taken into account physical processes by which water may be lost from the planets during or after their formation. 
When protoplanets migrate across the water snowline, their surface escape velocity is much larger than the thermal motions of a water vapour atmosphere (this is not true only for bodies smaller than a few hundred kilometres). We therefore do not expect protoplanets to lose water when they migrate past the water snowline. Collisions may present a bigger threat to water retention. We assume each collision between protoplanets and/or planetesimals leads to perfect merging, not taking into account the possibility of more destructive outcomes. According to  \citet{2018CeMDA.130....2B}, up to 75\%\ of the water can be lost in a hit-and-run collision between Mars-sized bodies (in a gas-free environment), and according to \citet{2010ApJ...719L..45M}, an entire icy mantle can be stripped off a rocky core by a violent giant impact. If violent collisions happen outside the snowline, a large fraction of the `lost' water may still end up in the planets via condensation on pebbles, but collisions interior to the snowline would lead to lower planet water fractions.

\citet{2017MNRAS.464.3728B} and more recently \citet{2017AJ....154..121B} have estimated the amount of water loss from the TRAPPIST-1 planets due to photodissociation as a result of high-energy stellar radiation after the disk disappeared. Extrapolating the current high-energy stellar irradiation back in time using an evolutionary model, \citet{2017AJ....154..121B} report an upper limit of tens of Earth oceans on the water loss of TRAPPIST-1g since its formation. The other planets have lost lesser amounts. Tens of Earth oceans corresponds to a few percent of an Earth mass, which is still smaller than the water contents of our simulated planets and of the present-day TRAPPIST-1 planets \citep{2018A&A...613A..68G,2018ApJ...865...20D}. 

If a significant fraction of the TRAPPIST-1 planets' water contents were lost during or post formation, a disk model in which planets form more water-richly than the currently observed TRAPPIST-1 planets may be more realistic (e.g. a smaller, and therefore less massive disk than our fiducial model, such as `model 10', see \tb{results}).

\subsection{Planetary systems other than TRAPPIST-1}
We have shown that compact systems of Earth-sized planets can form around M-dwarf stars under different disk conditions. The occurrence rate of small planets is higher around M-dwarfs than around FGK-stars \citep{2015ApJ...814..130M}, but more systems like TRAPPIST-1, with as many as seven small planets, have not been discovered yet. However, forming a large enough number of planets has proven to be no problem in our model, which leaves us wondering why no other compact `many-planet' systems have been discovered. One reason could be that a large fraction of compact multi-planet systems originally formed with more planets in a resonance chain, but became unstable \citep{2017MNRAS.470.1750I,2019arXiv190208694L,2019arXiv190208772I}.

A crucial feature of our model is that the icy pebble flux from the outer disk reaches the snowline such that the streaming instability outside the snowline can be triggered. If a massive planet outside the snowline were to open a gap and halt the pebble flux, planetesimal formation would shut down. We hypothesise that this is the main reason for the different architecture of compact multi-planet systems like TRAPPIST-1 (in which all planets have similar sizes) on the one hand, and a system like the solar system on the other. In the solar system, the early formation of Jupiter may have stalled icy planetesimal formation just exterior to the snowline, leaving only dry material available to form the rocky planets in the inner solar system \citep{2015Icar..258..418M,2019arXiv190208694L}.

\section{Conclusions}\label{sec:conclusions}
We have performed a numerical follow-up study regarding the formation model for compact planetary systems around M-dwarf stars, such as the TRAPPIST-1 system, which was presented in \citet{OLS2017}. We have coupled a Lagrangian dust evolution and planetesimal formation code \citep{2018A&A...620A.134S} to an N-body code adapted to account for gas drag, type-I migration, and pebble accretion \citep{2019arXiv190210062L}. This coupling enabled us, for the first time, to self-consistently model the planet formation process from small dust grains to full-sized planets, whilst keeping track of their water content. The main conclusions from this work can be summarised as follows:

\begin{enumerate}
    \item Our work has demonstrated that the planet formation scenario proposed in  \citet{OLS2017} indeed produces multiple Earth-sized planets (even without taking into account the pebble isolation mass), with water mass fractions on the order of 10\%. 
    \item In contrast to what was hypothesised in \citet{OLS2017}, we find that all planetesimals form in one, early streaming instability phase.
    \item Because the N-body code contains a stochastic component (the exact injection location of planetesimals), we find scatter in the characteristics of planetary systems (such as the number of planets, the average planet mass, and water fraction) resulting from simulations with the same input parameters. The amount of scatter is quantified in \tb{results}.
    \item Our planet formation model shows a universal trend in the water fraction with planet order: inner- and outermost planets are generally more water-rich than middle planets, independent of the model parameters. This `V-shape' in the variation of the planet water fraction with orbital distance emerges because the three different growth mechanisms (planetesimal accretion, wet pebble accretion, and dry pebble accretion) have different relative importance for the inner, middle, and outer planets. The importance of wet planetesimal accretion for planet growth diminishes with planet order, whereas pebble accretion becomes more important for planets with larger orbital distance. The effect is a decreasing water fraction with planet orbital distance. However, the ratio of wet pebble accretion to dry pebble accretion goes down with planet orbital distance, such that the water fraction goes up again for the outermost planets.   
   \end{enumerate}
\noindent
We reiterate that, due to computational limitations, in this work we have solely focused on the final masses and compositions of the planets, and have not followed the final dynamical configuration of planetary systems. In order to extend our formation model such that it also treats the long-term dynamics of the system, the N-body code should be extended into the very inner region of the disk (interior to $r_{\rm{domain}}$), or a dedicated dynamics study should be performed of this inner region, by making use of the results of the simulations presented in this paper.  

\begin{acknowledgements}
    D.S.\ and C.W.O\ are supported by the Netherlands Organization for Scientific Research (NWO; VIDI project 639.042.422). B.L.\ thanks the support of  the European Research Council (ERC Consolidator Grant 724687-PLANETESYS), and the Swedish Walter Gyllenberg Foundation. C.D.\ acknowledges the support of the Swiss National Foundation under grant PZ00P2\_174028 and the Swiss National Center for Competence in Research {\it PlanetS}. We thank the anonymous referee for his or her comments that helped improve the manuscript. 
\end{acknowledgements}

\appendix
\section{Mass-radius diagrams for multiple realisations of the fiducial model.}\label{multimr}
We attach as an appendix the mass-radius diagrams of the planets resulting from nine realisations of the fiducial simulation. 

\begin{sidewaysfigure*}[!t]
        \centering
                \includegraphics[width=0.33\textwidth]{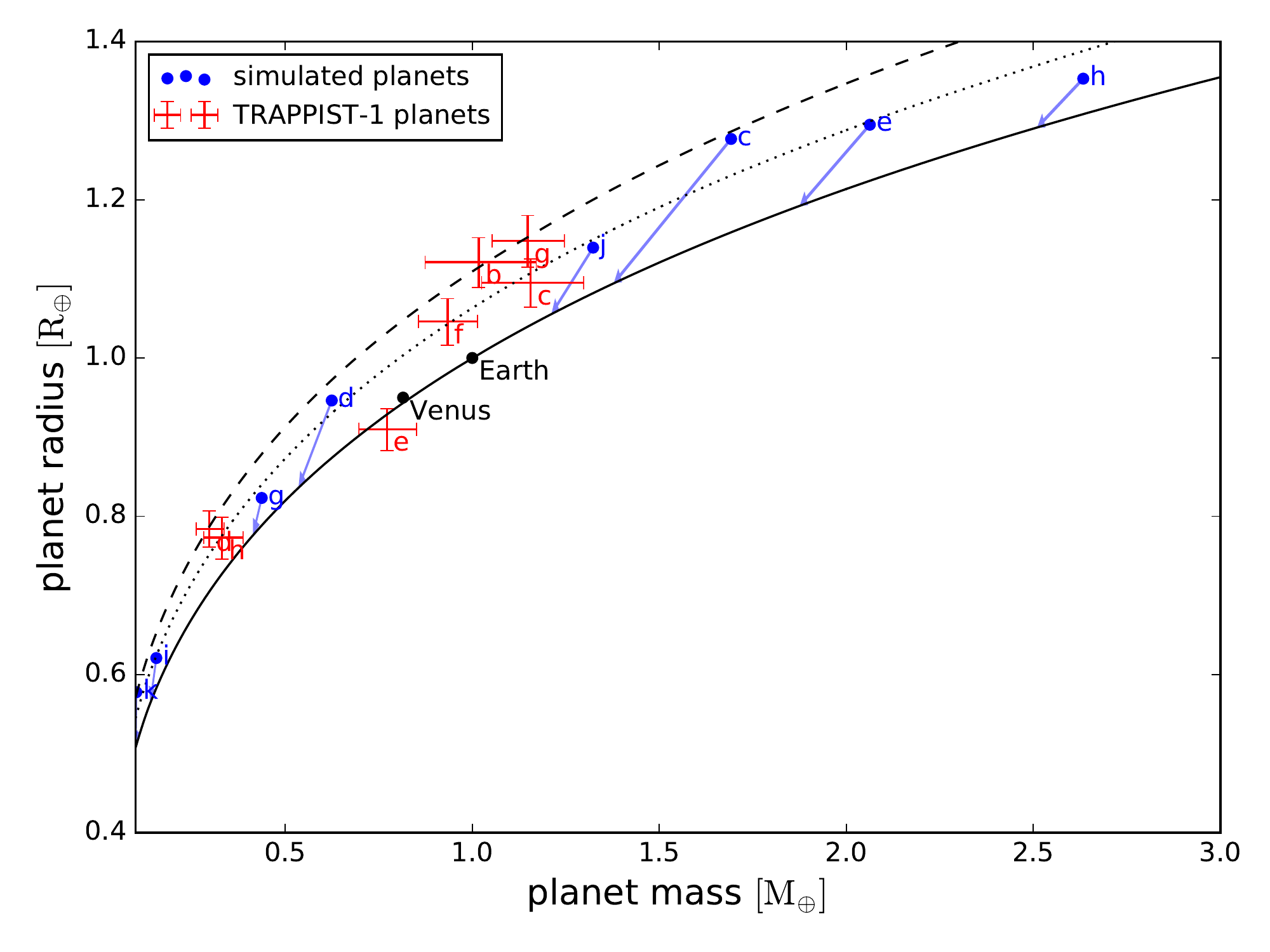}
                \includegraphics[width=0.33\textwidth]{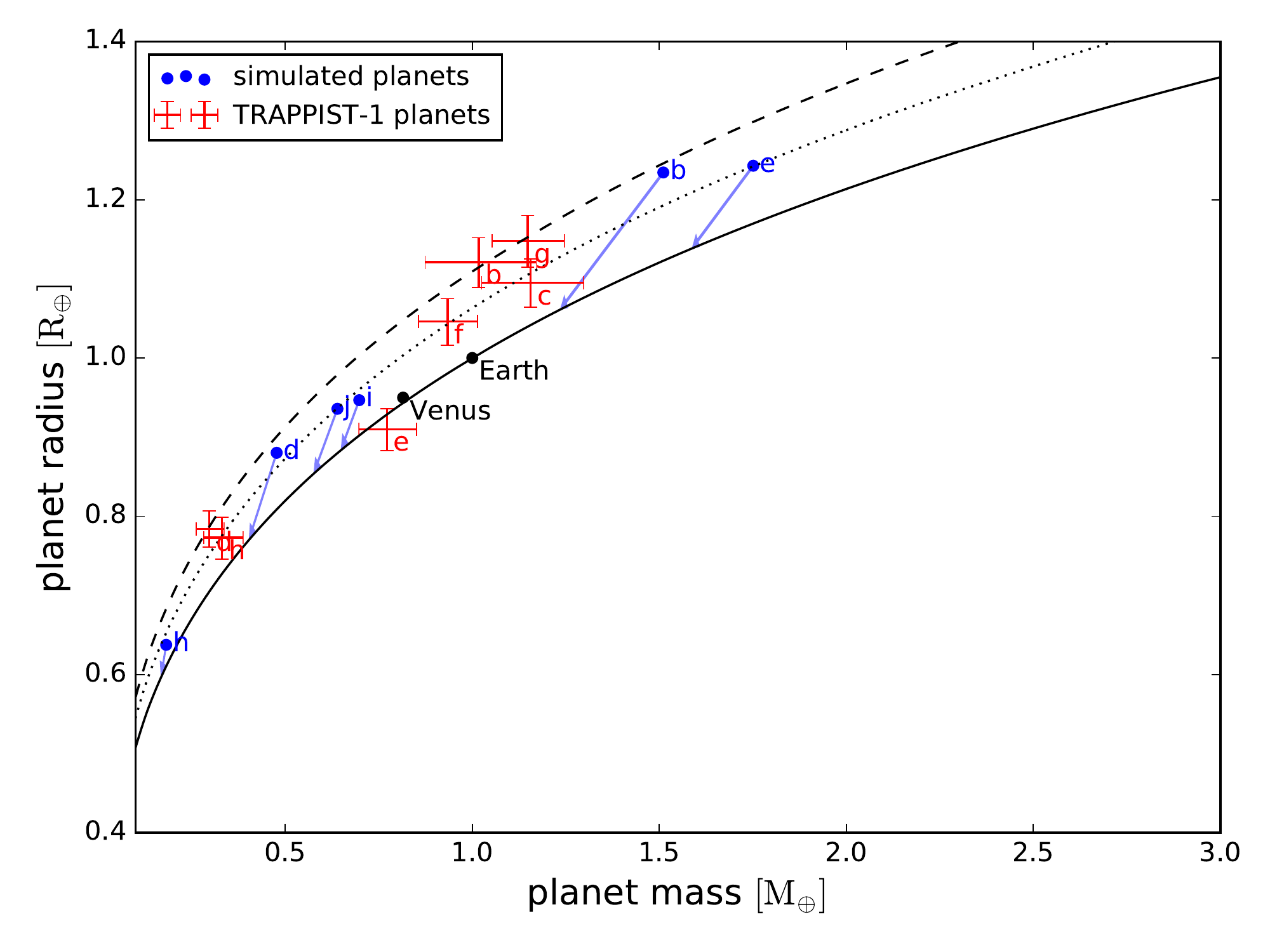}
                \includegraphics[width=0.33\textwidth]{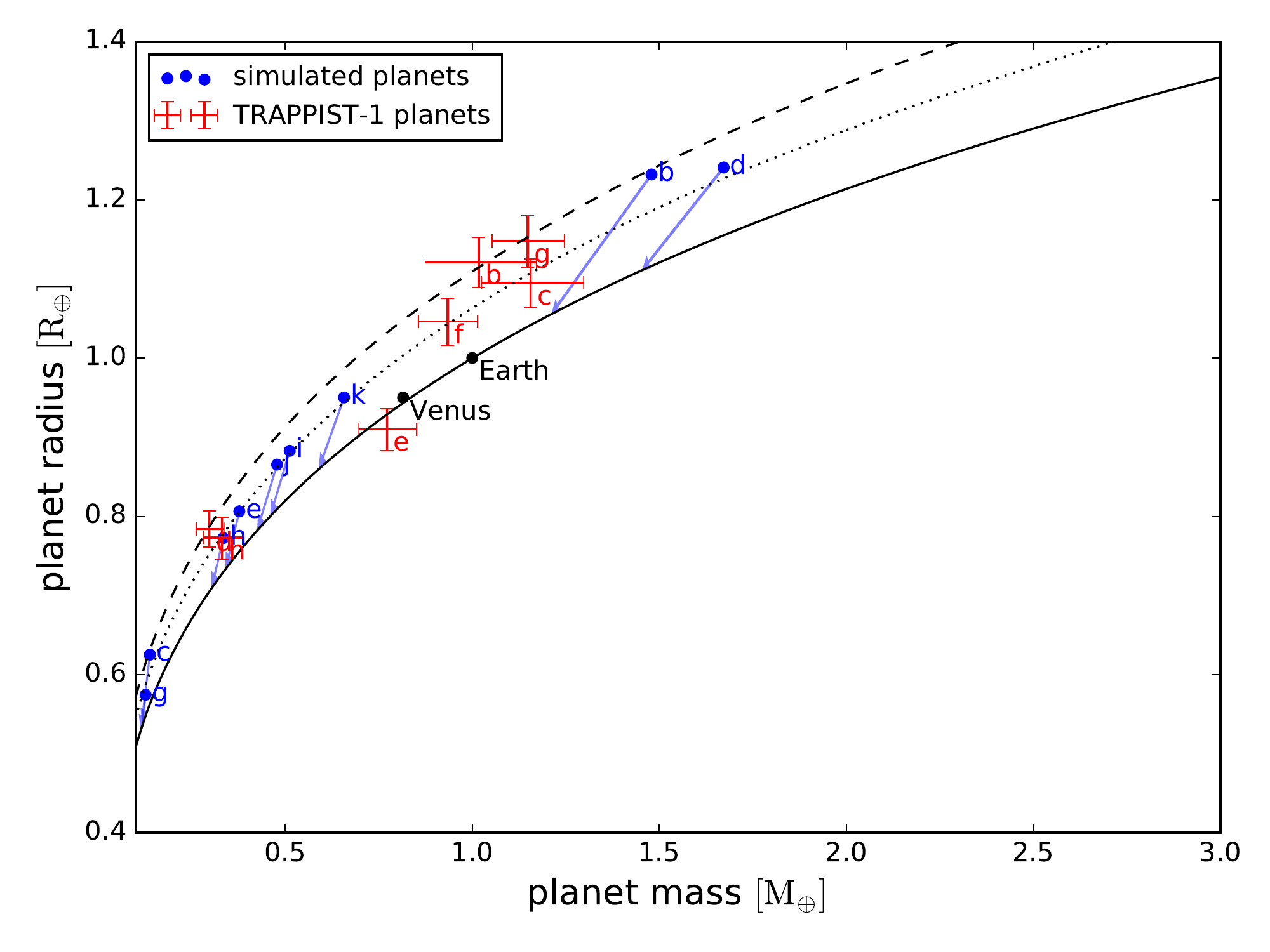}
                \includegraphics[width=0.33\textwidth]{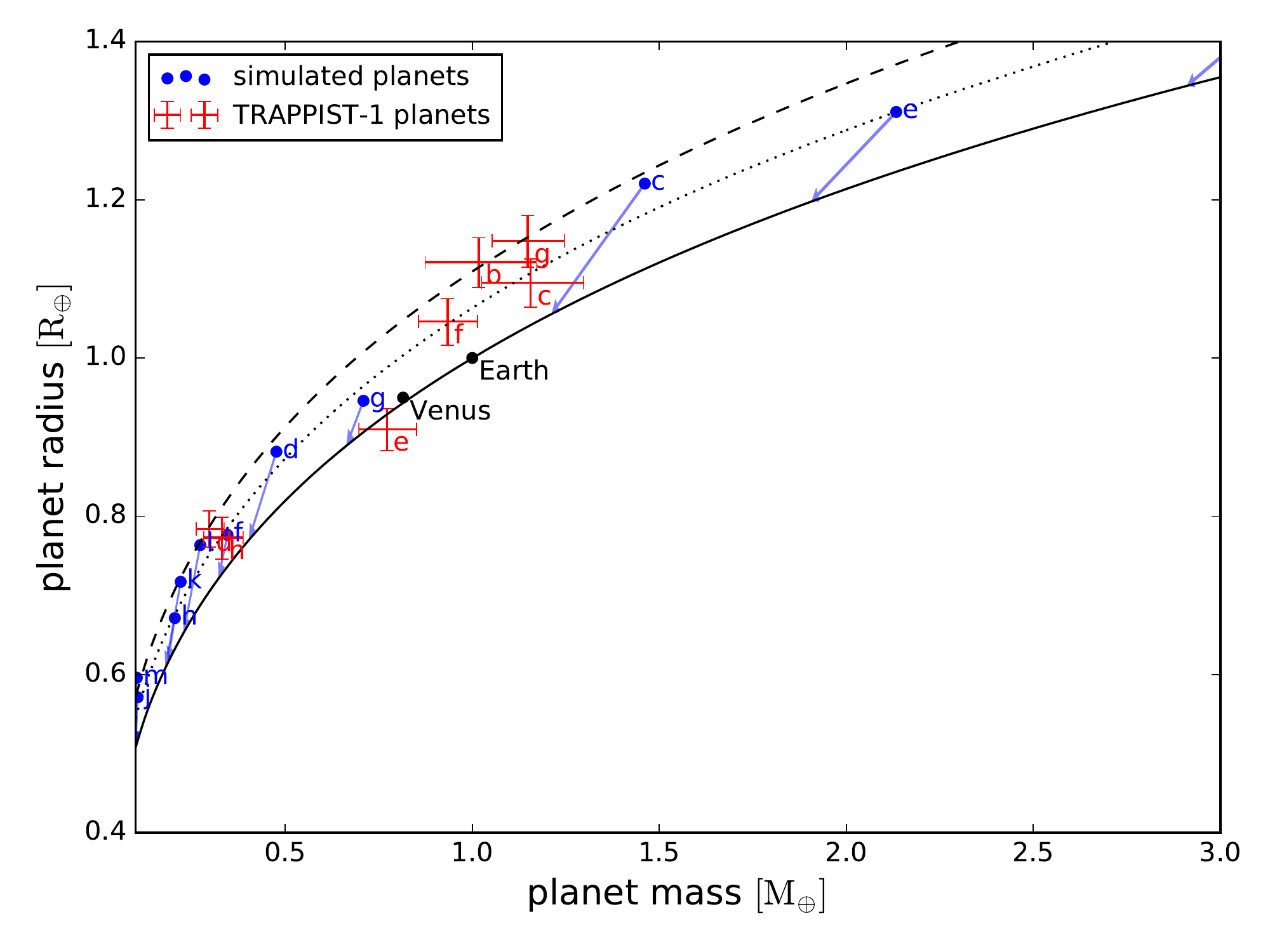}
                \includegraphics[width=0.33\textwidth]{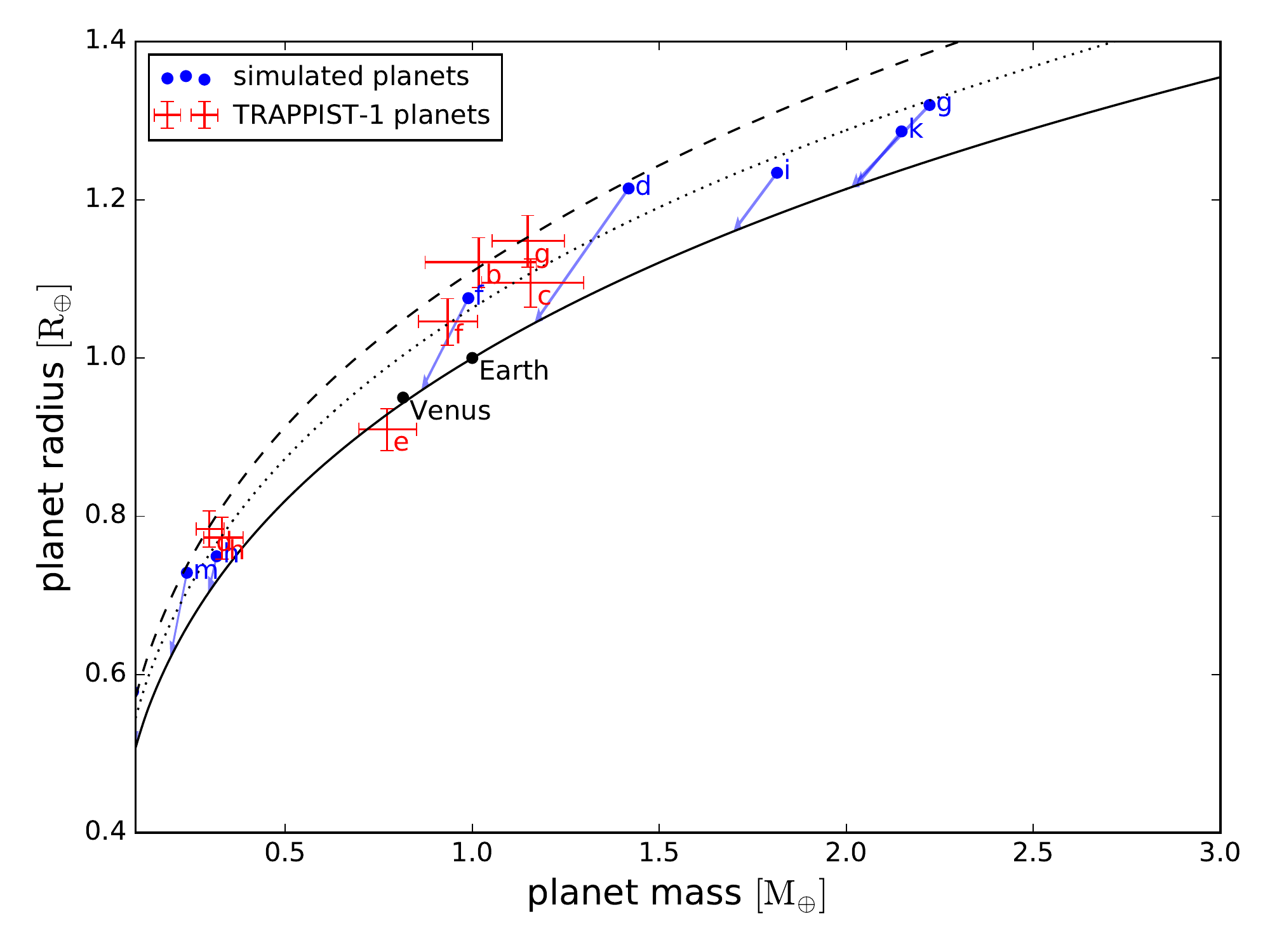}
                \includegraphics[width=0.33\textwidth]{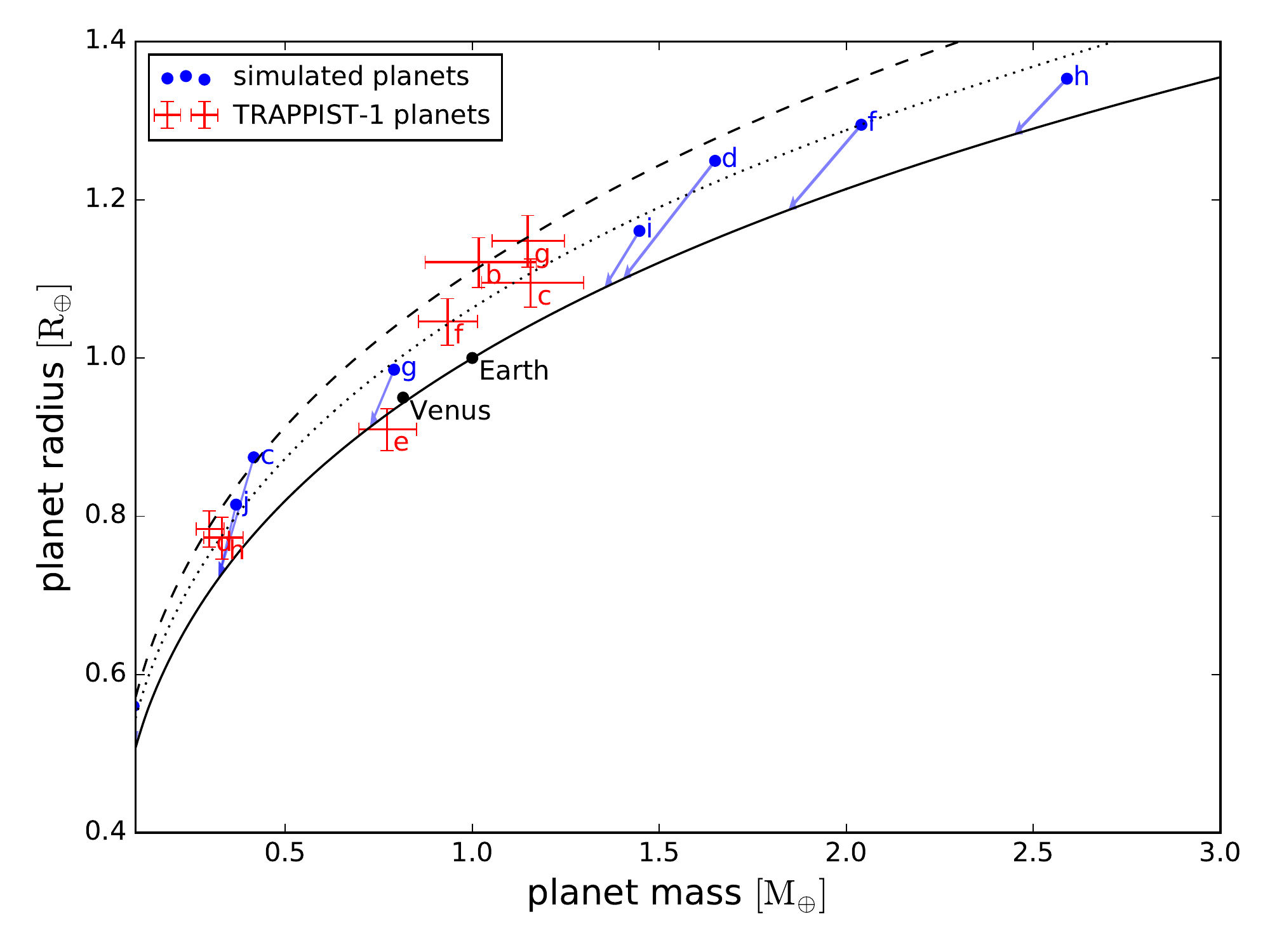}
                \includegraphics[width=0.33\textwidth]{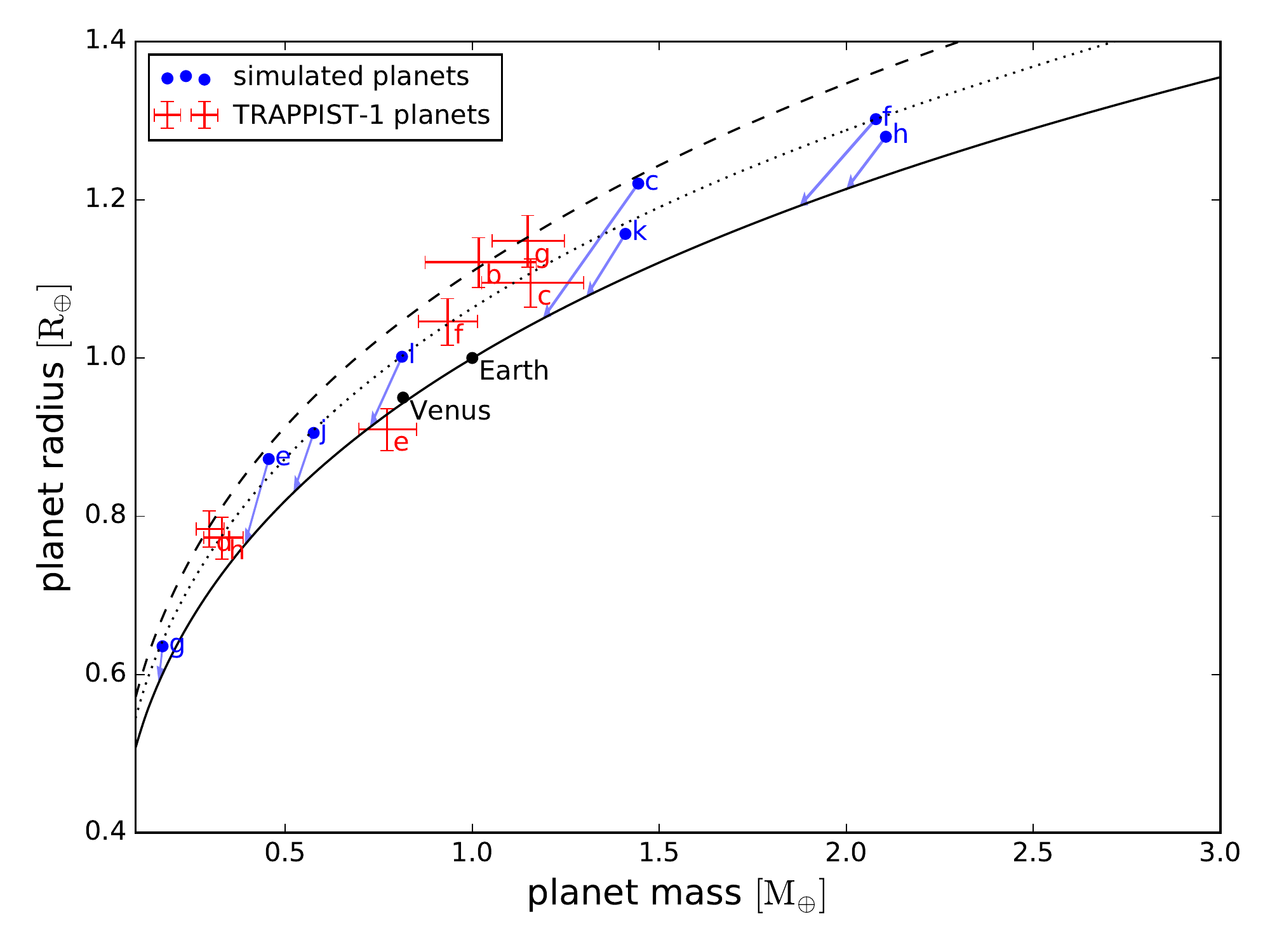}
                \includegraphics[width=0.33\textwidth]{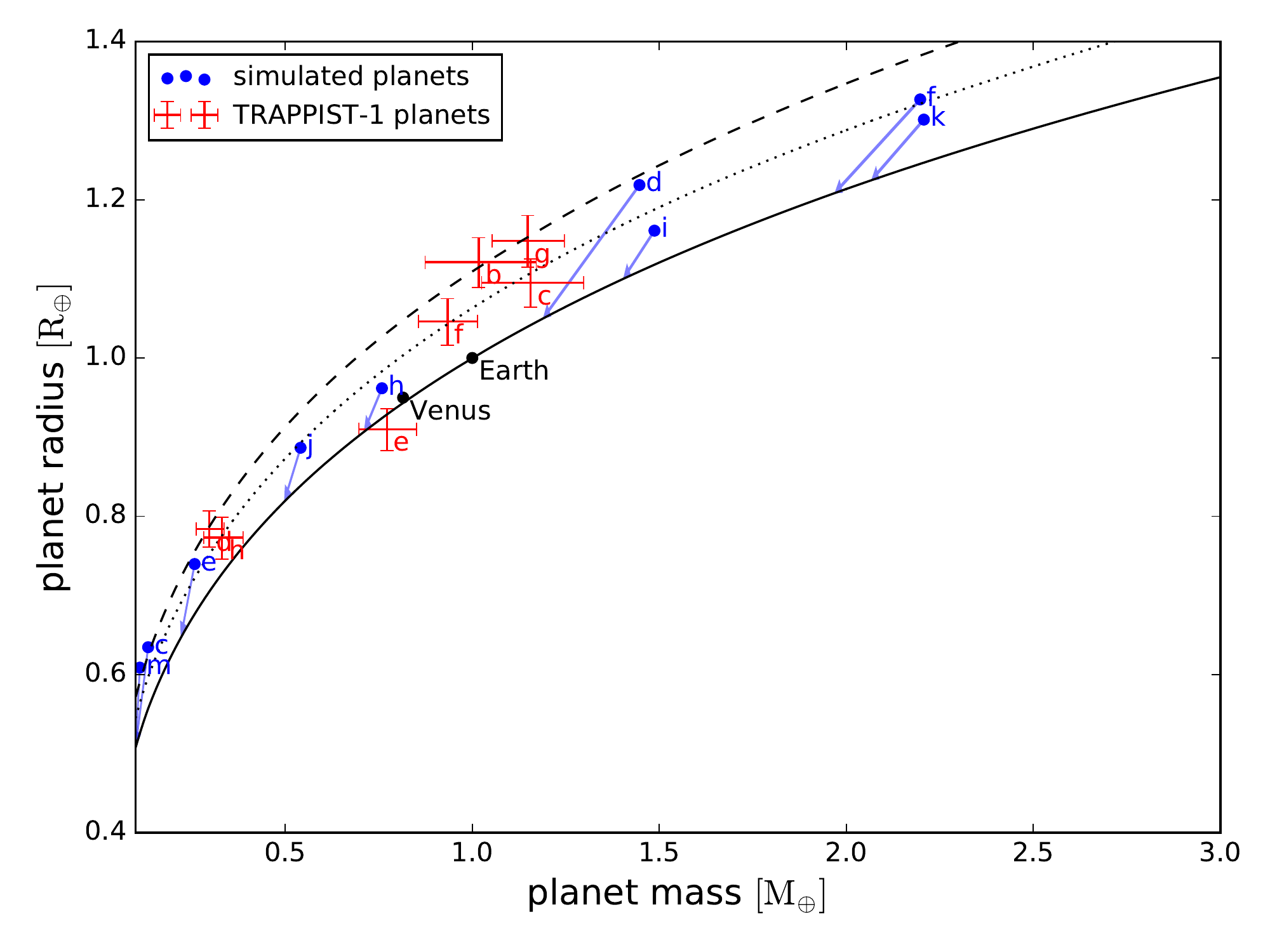}
                \includegraphics[width=0.33\textwidth]{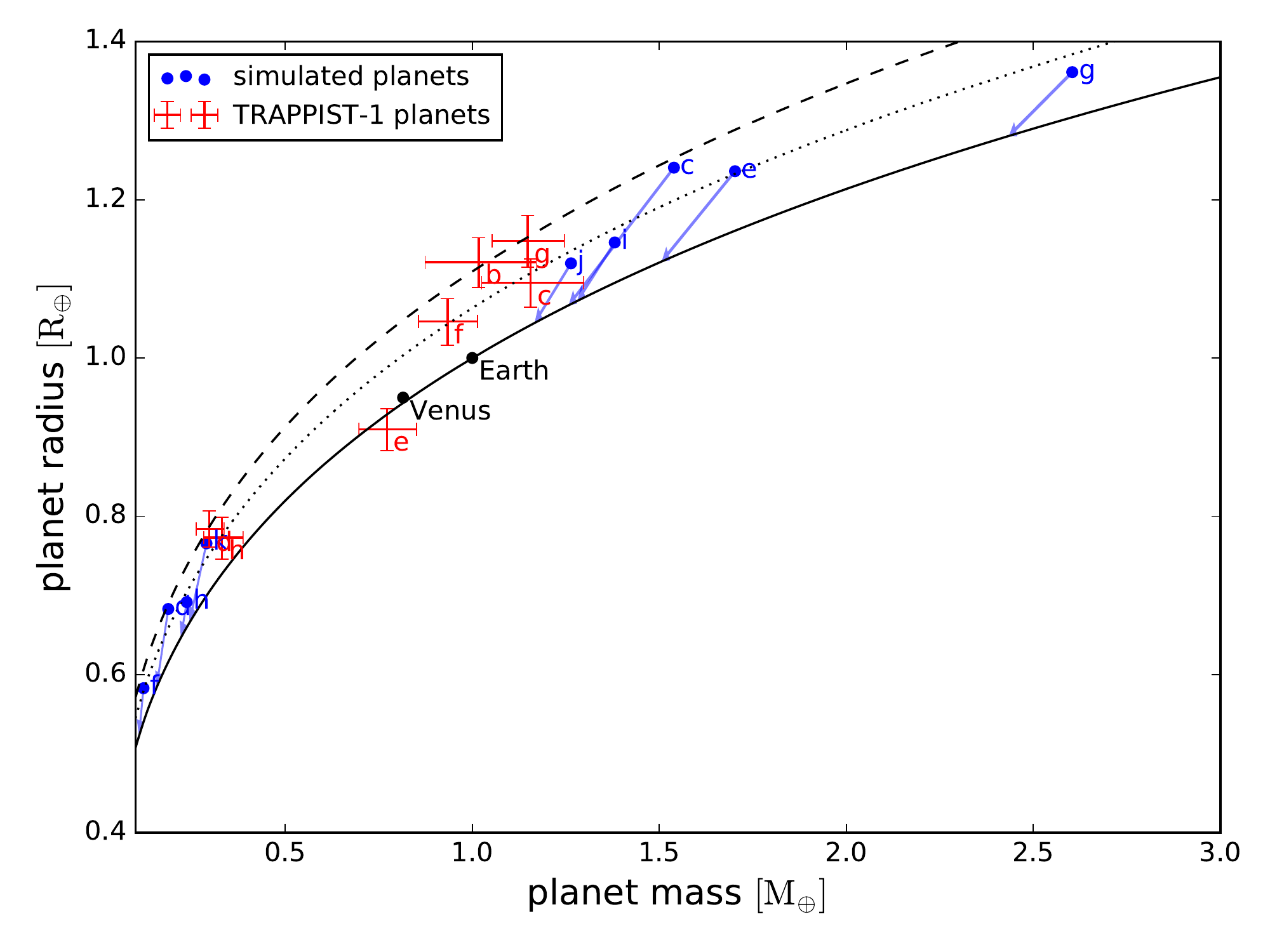}
                \caption{Mass-radius diagrams with simulated planets from nine different realisations of the fiducial model (parameter values listed in \tb{inputpar}). Simulated planets are plotted by blue dots; TRAPPIST-1 planets by red crosses (data taken from \citet{2018ApJ...865...20D}). The solid, dotted, and dashed lines correspond to mass-radius relations for a rocky composition and a water fraction of 0\%, 10\%, and 20\%, respectively. The mass-radius relations are the same as the ones used in \citet{2018ApJ...865...20D}. The arrows starting from the simulated planets depict the trajectories on the mass-radius diagram that the simulated planets would cover if they were to lose their entire water content.
\label{fig:ninetimes}}
\end{sidewaysfigure*}

\bibliographystyle{aa}
\bibliography{trappistbib}

\begin{thebibliography}{65}
\expandafter\ifx\csname natexlab\endcsname\relax\def\natexlab#1{#1}\fi

\bibitem[{{Abod} {et~al.}(2018){Abod}, {Simon}, {Li}, {Armitage}, {Youdin}, \&
  {Kretke}}]{2018arXiv181010018A}
{Abod}, C.~P., {Simon}, J.~B., {Li}, R., {et~al.} 2018, arXiv e-prints,
  arXiv:1810.10018

\bibitem[{{Alibert}(2019)}]{2019arXiv190109719A}
{Alibert}, Y. 2019, arXiv e-prints, arXiv:1901.09719

\bibitem[{{Ataiee} {et~al.}(2018){Ataiee}, {Baruteau}, {Alibert}, \&
  {Benz}}]{2018A&A...615A.110A}
{Ataiee}, S., {Baruteau}, C., {Alibert}, Y., \& {Benz}, W. 2018, \aap, 615,
  A110

\bibitem[{{Bitsch} {et~al.}(2018){Bitsch}, {Morbidelli}, {Johansen}, {Lega},
  {Lambrechts}, \& {Crida}}]{2018A&A...612A..30B}
{Bitsch}, B., {Morbidelli}, A., {Johansen}, A., {et~al.} 2018, \aap, 612, A30

\bibitem[{{Bitsch} {et~al.}(2014){Bitsch}, {Morbidelli}, {Lega}, {Kretke}, \&
  {Crida}}]{2014A&A...570A..75B}
{Bitsch}, B., {Morbidelli}, A., {Lega}, E., {Kretke}, K., \& {Crida}, A. 2014,
  \aap, 570, A75

\bibitem[{{Bitsch} {et~al.}(2019){Bitsch}, {Raymond}, \&
  {Izidoro}}]{2019A&A...624A.109B}
{Bitsch}, B., {Raymond}, S.~N., \& {Izidoro}, A. 2019, \aap, 624, A109

\bibitem[{{Blum} \& {M{\"u}nch}(1993)}]{1993Icar..106..151B}
{Blum}, J. \& {M{\"u}nch}, M. 1993, \icarus, 106, 151

\bibitem[{{Bolmont} {et~al.}(2017){Bolmont}, {Selsis}, {Owen}, {Ribas},
  {Raymond}, {Leconte}, \& {Gillon}}]{2017MNRAS.464.3728B}
{Bolmont}, E., {Selsis}, F., {Owen}, J.~E., {et~al.} 2017, \mnras, 464, 3728

\bibitem[{{Bourrier} {et~al.}(2017){Bourrier}, {de Wit}, {Bolmont},
  {Stamenkovi{\'c}}, {Wheatley}, {Burgasser}, {Delrez}, {Demory}, {Ehrenreich},
  {Gillon}, {Jehin}, {Leconte}, {Lederer}, {Lewis}, {Triaud}, \& {Van
  Grootel}}]{2017AJ....154..121B}
{Bourrier}, V., {de Wit}, J., {Bolmont}, E., {et~al.} 2017, \aj, 154, 121

\bibitem[{{Burger} {et~al.}(2018){Burger}, {Maindl}, \&
  {Sch{\"a}fer}}]{2018CeMDA.130....2B}
{Burger}, C., {Maindl}, T.~I., \& {Sch{\"a}fer}, C.~M. 2018, Celestial
  Mechanics and Dynamical Astronomy, 130, 2

\bibitem[{{Chambers}(1999)}]{1999MNRAS.304..793C}
{Chambers}, J.~E. 1999, \mnras, 304, 793

\bibitem[{{Ciesla} \& {Cuzzi}(2006)}]{2006Icar..181..178C}
{Ciesla}, F.~J. \& {Cuzzi}, J.~N. 2006, \icarus, 181, 178

\bibitem[{{Cridland} {et~al.}(2019){Cridland}, {Pudritz}, \&
  {Alessi}}]{2019MNRAS.484..345C}
{Cridland}, A.~J., {Pudritz}, R.~E., \& {Alessi}, M. 2019, \mnras, 484, 345

\bibitem[{{Delrez} {et~al.}(2018){Delrez}, {Gillon}, {Triaud}, {Demory}, {de
  Wit}, {Ingalls}, {Agol}, {Bolmont}, {Burdanov}, {Burgasser}, {Carey},
  {Jehin}, {Leconte}, {Lederer}, {Queloz}, {Selsis}, \& {Van
  Grootel}}]{2018MNRAS.475.3577D}
{Delrez}, L., {Gillon}, M., {Triaud}, A.~H.~M.~J., {et~al.} 2018, \mnras, 475,
  3577

\bibitem[{{Dorn} {et~al.}(2018){Dorn}, {Mosegaard}, {Grimm}, \&
  {Alibert}}]{2018ApJ...865...20D}
{Dorn}, C., {Mosegaard}, K., {Grimm}, S.~L., \& {Alibert}, Y. 2018, \apj, 865,
  20

\bibitem[{{Dr{\c a}{\.z}kowska} \& {Alibert}(2017)}]{2017A&A...608A..92D}
{Dr{\c a}{\.z}kowska}, J. \& {Alibert}, Y. 2017, \aap, 608, A92

\bibitem[{{Garaud} \& {Lin}(2007)}]{2007ApJ...654..606G}
{Garaud}, P. \& {Lin}, D.~N.~C. 2007, \apj, 654, 606

\bibitem[{{Gillon} {et~al.}(2016){Gillon}, {Jehin}, {Lederer}, {Delrez}, {de
  Wit}, {Burdanov}, {Van Grootel}, {Burgasser}, {Triaud}, {Opitom}, {Demory},
  {Sahu}, {Bardalez Gagliuffi}, {Magain}, \& {Queloz}}]{2016Natur.533..221G}
{Gillon}, M., {Jehin}, E., {Lederer}, S.~M., {et~al.} 2016, \nat, 533, 221

\bibitem[{{Gillon} {et~al.}(2017){Gillon}, {Triaud}, {Demory}, {Jehin}, {Agol},
  {Deck}, {Lederer}, {de Wit}, {Burdanov}, {Ingalls}, {Bolmont}, {Leconte},
  {Raymond}, {Selsis}, {Turbet}, {Barkaoui}, {Burgasser}, {Burleigh}, {Carey},
  {Chaushev}, {Copperwheat}, {Delrez}, {Fernand es}, {Holdsworth}, {Kotze},
  {Van Grootel}, {Almleaky}, {Benkhaldoun}, {Magain}, \&
  {Queloz}}]{2017Natur.542..456G}
{Gillon}, M., {Triaud}, A. H.~M.~J., {Demory}, B.-O., {et~al.} 2017, \nat, 542,
  456

\bibitem[{{Goldreich} \& {Schlichting}(2014)}]{2014AJ....147...32G}
{Goldreich}, P. \& {Schlichting}, H.~E. 2014, \aj, 147, 32

\bibitem[{{Grimm} {et~al.}(2018){Grimm}, {Demory}, {Gillon}, {Dorn}, {Agol},
  {Burdanov}, {Delrez}, {Sestovic}, {Triaud}, {Turbet}, {Bolmont}, {Caldas},
  {Wit}, {Jehin}, {Leconte}, {Raymond}, {Grootel}, {Burgasser}, {Carey},
  {Fabrycky}, {Heng}, {Hernandez}, {Ingalls}, {Lederer}, {Selsis}, \&
  {Queloz}}]{2018A&A...613A..68G}
{Grimm}, S.~L., {Demory}, B.-O., {Gillon}, M., {et~al.} 2018, \aap, 613, A68

\bibitem[{{Gundlach} \& {Blum}(2015)}]{2015ApJ...798...34G}
{Gundlach}, B. \& {Blum}, J. 2015, \apj, 798, 34

\bibitem[{{Hendler} {et~al.}(2017){Hendler}, {Mulders}, {Pascucci},
  {Greenwood}, {Kamp}, {Henning}, {M{\'e}nard}, {Dent}, \&
  {Evans}}]{2017ApJ...841..116H}
{Hendler}, N.~P., {Mulders}, G.~D., {Pascucci}, I., {et~al.} 2017, \apj, 841,
  116

\bibitem[{{Izidoro} {et~al.}(2019){Izidoro}, {Bitsch}, {Raymond}, {Johansen},
  {Morbidelli}, {Lambrechts}, \& {Jacobson}}]{2019arXiv190208772I}
{Izidoro}, A., {Bitsch}, B., {Raymond}, S.~N., {et~al.} 2019, arXiv e-prints,
  arXiv:1902.08772

\bibitem[{{Izidoro} {et~al.}(2017){Izidoro}, {Ogihara}, {Raymond},
  {Morbidelli}, {Pierens}, {Bitsch}, {Cossou}, \&
  {Hersant}}]{2017MNRAS.470.1750I}
{Izidoro}, A., {Ogihara}, M., {Raymond}, S.~N., {et~al.} 2017, \mnras, 470,
  1750

\bibitem[{{Johansen} {et~al.}(2015){Johansen}, {Mac Low}, {Lacerda}, \&
  {Bizzarro}}]{2015SciA....1E0109J}
{Johansen}, A., {Mac Low}, M.-M., {Lacerda}, P., \& {Bizzarro}, M. 2015,
  Science Advances, 1, 1500109

\bibitem[{{Kretke} \& {Lin}(2007)}]{2007ApJ...664L..55K}
{Kretke}, K.~A. \& {Lin}, D.~N.~C. 2007, \apj, 664, L55

\bibitem[{{Krijt} {et~al.}(2016){Krijt}, {Ormel}, {Dominik}, \&
  {Tielens}}]{2016A&A...586A..20K}
{Krijt}, S., {Ormel}, C.~W., {Dominik}, C., \& {Tielens}, A.~G.~G.~M. 2016,
  \aap, 586, A20

\bibitem[{{Lambrechts} {et~al.}(2014){Lambrechts}, {Johansen}, \&
  {Morbidelli}}]{2014A&A...572A..35L}
{Lambrechts}, M., {Johansen}, A., \& {Morbidelli}, A. 2014, \aap, 572, A35

\bibitem[{{Lambrechts} {et~al.}(2019){Lambrechts}, {Morbidelli}, {Jacobson},
  {Johansen}, {Bitsch}, {Izidoro}, \& {Raymond}}]{2019arXiv190208694L}
{Lambrechts}, M., {Morbidelli}, A., {Jacobson}, S.~A., {et~al.} 2019, arXiv
  e-prints, arXiv:1902.08694

\bibitem[{{Li} {et~al.}(2018){Li}, {Youdin}, \& {Simon}}]{2018ApJ...862...14L}
{Li}, R., {Youdin}, A.~N., \& {Simon}, J.~B. 2018, \apj, 862, 14

\bibitem[{{Liu} \& {Ormel}(2018)}]{2018A&A...615A.138L}
{Liu}, B. \& {Ormel}, C.~W. 2018, \aap, 615, A138

\bibitem[{{Liu} {et~al.}(2019){Liu}, {Ormel}, \&
  {Johansen}}]{2019arXiv190210062L}
{Liu}, B., {Ormel}, C.~W., \& {Johansen}, A. 2019, arXiv e-prints,
  arXiv:1902.10062

\bibitem[{{Liu} {et~al.}(2017){Liu}, {Ormel}, \& {Lin}}]{2017A&A...601A..15L}
{Liu}, B., {Ormel}, C.~W., \& {Lin}, D. N.~C. 2017, \aap, 601, A15

\bibitem[{{Luger} {et~al.}(2017){Luger}, {Sestovic}, {Kruse}, {Grimm},
  {Demory}, {Agol}, {Bolmont}, {Fabrycky}, {Fernandes}, {Van Grootel},
  {Burgasser}, {Gillon}, {Ingalls}, {Jehin}, {Raymond}, {Selsis}, {Triaud},
  {Barclay}, {Barentsen}, {Howell}, {Delrez}, {de Wit}, {Foreman-Mackey},
  {Holdsworth}, {Leconte}, {Lederer}, {Turbet}, {Almleaky}, {Benkhaldoun},
  {Magain}, {Morris}, {Heng}, \& {Queloz}}]{2017NatAs...1E.129L}
{Luger}, R., {Sestovic}, M., {Kruse}, E., {et~al.} 2017, Nature Astronomy, 1,
  0129

\bibitem[{{Lynden-Bell} \& {Pringle}(1974)}]{1974MNRAS.168..603L}
{Lynden-Bell}, D. \& {Pringle}, J.~E. 1974, \mnras, 168, 603

\bibitem[{{Manara} {et~al.}(2017){Manara}, {Testi}, {Herczeg}, {Pascucci},
  {Alcal{\'a}}, {Natta}, {Antoniucci}, {Fedele}, {Mulders}, {Henning},
  {Mohanty}, {Prusti}, \& {Rigliaco}}]{2017A&A...604A.127M}
{Manara}, C.~F., {Testi}, L., {Herczeg}, G.~J., {et~al.} 2017, \aap, 604, A127

\bibitem[{{Marcus} {et~al.}(2010){Marcus}, {Sasselov}, {Stewart}, \&
  {Hernquist}}]{2010ApJ...719L..45M}
{Marcus}, R.~A., {Sasselov}, D., {Stewart}, S.~T., \& {Hernquist}, L. 2010,
  \apj, 719, L45

\bibitem[{{Martin} \& {Livio}(2012)}]{2012MNRAS.425L...6M}
{Martin}, R.~G. \& {Livio}, M. 2012, \mnras, 425, L6

\bibitem[{{Morbidelli} {et~al.}(2015){Morbidelli}, {Lambrechts}, {Jacobson}, \&
  {Bitsch}}]{2015Icar..258..418M}
{Morbidelli}, A., {Lambrechts}, M., {Jacobson}, S., \& {Bitsch}, B. 2015,
  \icarus, 258, 418

\bibitem[{{Morbidelli} \& {Nesvorny}(2012)}]{2012A&A...546A..18M}
{Morbidelli}, A. \& {Nesvorny}, D. 2012, \aap, 546, A18

\bibitem[{{Mulders} {et~al.}(2015){Mulders}, {Pascucci}, \&
  {Apai}}]{2015ApJ...814..130M}
{Mulders}, G.~D., {Pascucci}, I., \& {Apai}, D. 2015, \apj, 814, 130

\bibitem[{{Musiolik} \& {Wurm}(2019)}]{2019ApJ...873...58M}
{Musiolik}, G. \& {Wurm}, G. 2019, \apj, 873, 58

\bibitem[{{Nakagawa} {et~al.}(1986){Nakagawa}, {Sekiya}, \&
  {Hayashi}}]{NakagawaEtal1986}
{Nakagawa}, Y., {Sekiya}, M., \& {Hayashi}, C. 1986, Icarus, 67, 375

\bibitem[{{Oka} {et~al.}(2011){Oka}, {Nakamoto}, \&
  {Ida}}]{2011ApJ...738..141O}
{Oka}, A., {Nakamoto}, T., \& {Ida}, S. 2011, \apj, 738, 141

\bibitem[{{Ormel} \& {Liu}(2018)}]{2018A&A...615A.178O}
{Ormel}, C.~W. \& {Liu}, B. 2018, \aap, 615, A178

\bibitem[{{Ormel} {et~al.}(2017){Ormel}, {Liu}, \& {Schoonenberg}}]{OLS2017}
{Ormel}, C.~W., {Liu}, B., \& {Schoonenberg}, D. 2017, \aap, 604, A1

\bibitem[{{Papaloizou} {et~al.}(2018){Papaloizou}, {Szuszkiewicz}, \&
  {Terquem}}]{2018MNRAS.476.5032P}
{Papaloizou}, J.~C.~B., {Szuszkiewicz}, E., \& {Terquem}, C. 2018, \mnras, 476,
  5032

\bibitem[{{Piso} {et~al.}(2015){Piso}, {{\"O}berg}, {Birnstiel}, \&
  {Murray-Clay}}]{2015arXiv151105563P}
{Piso}, A.-M.~A., {{\"O}berg}, K.~I., {Birnstiel}, T., \& {Murray-Clay}, R.~A.
  2015, \apj, 815, 109

\bibitem[{{Ricci} {et~al.}(2013){Ricci}, {Isella}, {Carpenter}, \&
  {Testi}}]{2013ApJ...764L..27R}
{Ricci}, L., {Isella}, A., {Carpenter}, J.~M., \& {Testi}, L. 2013, \apj, 764,
  L27

\bibitem[{{Ros} \& {Johansen}(2013)}]{2013A&A...552A.137R}
{Ros}, K. \& {Johansen}, A. 2013, \aap, 552, A137

\bibitem[{{Rosotti} {et~al.}(2019){Rosotti}, {Tazzari}, {Booth}, {Testi},
  {Lodato}, \& {Clarke}}]{2019MNRAS.486.4829R}
{Rosotti}, G.~P., {Tazzari}, M., {Booth}, R.~A., {et~al.} 2019, \mnras, 486,
  4829

\bibitem[{{Sch{\"a}fer} {et~al.}(2017){Sch{\"a}fer}, {Yang}, \&
  {Johansen}}]{2017A&A...597A..69S}
{Sch{\"a}fer}, U., {Yang}, C.-C., \& {Johansen}, A. 2017, \aap, 597, A69

\bibitem[{{Schoonenberg} \& {Ormel}(2017)}]{SO2017}
{Schoonenberg}, D. \& {Ormel}, C.~W. 2017, \aap, 602, A21

\bibitem[{{Schoonenberg} {et~al.}(2018){Schoonenberg}, {Ormel}, \&
  {Krijt}}]{2018A&A...620A.134S}
{Schoonenberg}, D., {Ormel}, C.~W., \& {Krijt}, S. 2018, \aap, 620, A134

\bibitem[{{Shakura} \& {Sunyaev}(1973)}]{1973A&A....24..337S}
{Shakura}, N.~I. \& {Sunyaev}, R.~A. 1973, \aap, 24, 337

\bibitem[{{Simon} {et~al.}(2016){Simon}, {Armitage}, {Li}, \&
  {Youdin}}]{2016ApJ...822...55S}
{Simon}, J.~B., {Armitage}, P.~J., {Li}, R., \& {Youdin}, A.~N. 2016, \apj,
  822, 55

\bibitem[{{Sirono}(1999)}]{1999A&A...347..720S}
{Sirono}, S. 1999, \aap, 347, 720

\bibitem[{{Stevenson} \& {Lunine}(1988)}]{1988Icar...75..146S}
{Stevenson}, D.~J. \& {Lunine}, J.~I. 1988, \icarus, 75, 146

\bibitem[{{Tanaka} {et~al.}(2002){Tanaka}, {Takeuchi}, \&
  {Ward}}]{2002ApJ...565.1257T}
{Tanaka}, H., {Takeuchi}, T., \& {Ward}, W.~R. 2002, \apj, 565, 1257

\bibitem[{{Teyssandier} \& {Terquem}(2014)}]{2014MNRAS.443..568T}
{Teyssandier}, J. \& {Terquem}, C. 2014, \mnras, 443, 568

\bibitem[{{Unterborn} {et~al.}(2018{\natexlab{a}}){Unterborn}, {Desch},
  {Hinkel}, \& {Lorenzo}}]{2018NatAs...2..297U}
{Unterborn}, C.~T., {Desch}, S.~J., {Hinkel}, N.~R., \& {Lorenzo}, A.
  2018{\natexlab{a}}, Nature Astronomy, 2, 297

\bibitem[{{Unterborn} {et~al.}(2018{\natexlab{b}}){Unterborn}, {Hinkel}, \&
  {Desch}}]{2018RNAAS...2c.116U}
{Unterborn}, C.~T., {Hinkel}, N.~R., \& {Desch}, S.~J. 2018{\natexlab{b}},
  Research Notes of the American Astronomical Society, 2, 116

\bibitem[{{Yang} \& {Johansen}(2016)}]{2016ApJS..224...39Y}
{Yang}, C.-C. \& {Johansen}, A. 2016, The Astrophysical Journal Supplement
  Series, 224, 39

\bibitem[{{Youdin} \& {Lithwick}(2007)}]{2007Icar..192..588Y}
{Youdin}, A.~N. \& {Lithwick}, Y. 2007, \icarus, 192, 588

\end{thebibliography}

\end{document}